\documentclass[12pt]{JHEP3}
\usepackage{amssymb,amsfonts}
\usepackage{amsmath}
\usepackage{graphicx}
\usepackage{amsmath,epsfig}
\usepackage{amssymb,amsfonts}
\usepackage{latexsym}
\def\hri#1#2{\href{http://arxiv.org/abs/#1}{[ArXiv:#1]#2}}
\def\hre#1#2{\href{http://arxiv.org/abs/#1/#2}{[ArXiv:#1/#2]}}
\def\hspi#1#2{\href{http://www.slac.stanford.edu/spires/find/hep/www?irn=#1}{#2}}

\def\s{\sigma}
\def\be{\begin{equation}}
\def\ee{\end{equation}}
\def\bea{\begin{eqnarray}}
\def\eea{\end{eqnarray}}
\def\sp{\;\;\;,\;\;\;}
\def\l{\lambda}

\def\lab{\label}
\def\f{\phi}
\def\z{\zeta}
\def\o{\omega}
\def\le{\left}
\def\ri{\right}

\def\half{\frac12}

\def\cO{{\cal O}}
\def\m{\mu}
\def\n{\nu}
\def\td{\tilde}
\def\d{\delta}
\def\6{\partial}
\def\del{\nabla}
\def\ls{\ell_s}
\def\a{\alpha}
\def\b{\beta}

\title{Thermal Transport and Drag Force in  Improved Holographic QCD}

\author{Umut G\"ursoy$^1$, Elias Kiritsis$^{2}$, Georgios Michalogiorgakis$^3$ and Francesco Nitti$^4$\\
$^1$\href{http://www1.phys.uu.nl/wwwitf}{Institute for Theoretical Physics, Utrecht University;
Leuvenlaan 4, 3584 CE Utrecht, The Netherlands.}\\
~\\
$^2$\href{http://hep.physics.uoc.gr/}{Department of Physics, University of Crete
71003 Heraklion, Greece}\\
~\\
$^3$\href{http://cpht.polytechnique.fr/cpht/cordes/}{CPHT, Ecole Polytechnique, CNRS,
 91128, Palaiseau, France}\\
 (UMR du CNRS 7644)\\
~\\
$^4$\href{http://www.apc.univ-paris7.fr}{APC, Universit\'e Paris 7, \\ B\^atiment Condorcet, F-75205, Paris Cedex 13, France}\\
 (UMR du CNRS 7164).

}

\preprint{arXiv:0906.1890 [hep-ph]\\CPHT-RR052.0609\\ ITP-UU-09/22\\ SPIN-09/21\\CCTP-2009-18}

\abstract{We calculate the bulk viscosity, drag force and jet quenching parameter in Improved Holographic QCD.
We find that the bulk viscosity rises near the phase transition but does not exceed the shear viscosity.
The drag force shows the effects of asymptotic freedom both as a function of velocity and temperature.
It indicates diffusion times of heavy quarks in rough agreement with data. The jet quenching parameter values computed via the light-like Wilson loop
are in the lower
range suggested by data.  }

\begin{document}
\maketitle

\section{Introduction}\label{INTRO}

A novel window in the physics of the strong interactions has been provided recently by the experimental efforts at RHIC, \cite{rhic} .
The consensus on the existing data is that shortly after the collision, a ball of quark-gluon plasma (QGP) forms that is at thermal equilibrium,
and  subsequently expands until its temperature falls below the QCD transition (or crossover) where it finally hadronizes.
Relativistic hydrodynamics describes very well the QGP  \cite{lr,lr2}, with a shear-viscosity to entropy density ratio close to
the universal value suggested by the holographic formulation of  ${\cal N}=4$ SYM, \cite{pss}.

The QGP is at strong coupling, and it necessitates a treatment beyond perturbative QCD approaches, \cite{review}.
There are several observables that seem to be important in understanding measured features of the collisions.
They translate into transport properties of the strongly coupled plasma, and reliable methods for the calculation are in need.

A fist class of transport coefficients are viscosity coefficients.\footnote{These are the leading transport coefficients in the derivative expansion.
There are subleading coefficients that have been calculated recently for ${\cal N}=4$ SYM, \cite{sub}. However, at the present level of accuracy,
 they cannot affect substantially the comparison to experimental data, \cite{lr}.}
A general fluid is characterized by two viscosity coefficients, the shear $\eta$ and the bulk viscosity $\zeta$.
The shear viscosity in strongly coupled theories described by gravity duals was shown to be universal, \cite{pss}.
In particular, the ratio $\eta/s$, with $s$ the entropy density, is equal to ${1\over 4\pi}$.
This is correlated to the universality of low-energy scattering of gravitons from black-holes.
It is also known that deviations from this value can only be generated by higher curvature terms that contain the Riemann tensor
(as opposed to the Ricci tensor of the scalar curvature).
In QCD, as the theory is strongly coupled in the temperature range $T_c\leq T\leq 3T_c$,
we would expect that $\eta/s\simeq {1\over 4\pi}$.
Recent lattice calculations, \cite{mshear} agree with this expectations although potential systematic
 errors in lattice calculations of transport coefficients can be large.

Conformal invariance forces the bulk viscosity to vanish. Therefore the  ${\cal N}=4$ SYM plasma, being a conformal fluid, has vanishing bulk viscosity.
QCD on the other hand is not a conformal theory. The classical theory is however conformally invariant and asymptotic freedom implies that
conformal invariance is a good approximation in the UV.
 This would suggest that the bulk viscosity is negligible at large temperatures.
   However it is not expected to be so in the IR: as mentioned earlier lattice data indicate that in the relevant
RHIC range  $1\leq {T\over T_c}\leq 3$ the QGP
seems not to be a fully  conformal fluid.
 Therefore the bulk viscosity may play a role near the phase transition.

 So far there have been two approaches that have calculated the bulk viscosity
 in YM/QCD, \cite{Viscos1,Viscos2,sum,Viscos3} and have both indicated that the bulk viscosity
 rises near the phase transition as naive expectation would suggest.
 The first used the method of sum rules in conjunction with input from Lattice thermodynamics, \cite{Viscos1,Viscos2,sum}.
 It suggested a dramatic rise of the bulk viscosity near $T_c$ although the absolute normalization of the result is uncertain.
 The reason is that this method relies on an ansatz for the density associated with stress-tensor two point functions that are otherwise unknown.

 The second method \cite{Viscos3} relies on a direct computation of the density at low frequency of the appropriate stress-tensor two-point function.
 As this computation is necessarily Euclidean, an analytic continuation is necessary.  The values at a finite number of discrete Matsubara frequencies
 are not enough to analytically continue. An ansatz for the continuous density is also used here, which presents again a potentially
 large systematic uncertainty.

 We will see in the present work that our findings support a rise of the bulk viscosity near $T_c$, but the values are much smaller than previously expected.
 Studies of how this affects hydrodynamics at RHIC, \cite{heinz} suggest that this implies a small fall in radial and elliptic flow.

Another class of interesting experimental observables is associated with quarks, and comes under
the label of ``jet quenching". Central to this is the expectation that an energetic quark will loose energy very fast in the quark-gluon plasma
because of strong coupling. This has as a side effect that back-to back jets are suppressed. Moreover if a pair of energetic quarks is generated
near the plasma boundary then one will exit fast the plasma and register as an energetic jet, while the other will thermalize and its
identity will disappear. This has been clearly observed at RHIC and used to study the energy loss of quarks in the quark-gluon plasma.

Heavy quarks are of extra importance, as their mass masks some low-energy strong interaction effects, and can be therefore cleaner probes of plasma energy loss.
There are important electron observables at RHIC, \cite{phenix} that can probe heavy-quark energy loss in the strongly coupled quark-gluon plasma.
Such observables are also expected to play an important role in LHC \cite{urs}.

A perturbative QCD approach to calculate the energy loss of a heavy quark in the plasma has been pursued by calculating radiative energy loss, \cite{rel}.
However its application to the RHIC plasma has recently raised  problems,
 based on comparison with data. A phenomenological coefficient used in such cases is known as the jet quenching coefficient
$\hat q$, and is defined as the rate of change of the average value of transverse momentum square of a probe.
Current fits,  \cite{phenix,lan}, indicate that a value of order 10 $GeV^2/fm$ or more is needed to describe the data
 while perturbative approaches are trustworthy at much lower values.

Several attempts were made to compute quark energy loss in the holographic context, relevant for   ${\cal N}=4$ SYM\footnote{Most are reviewed in
\cite{gubser-review}.}. In some of them \cite{Liu1,Liu2} the jet-quenching coefficient $\hat q$ was
calculated via its relationship to a light-like Wilson loop.
Holography was then used to calculate the appropriate Wilson loop.
   The $\hat q$ obtained scales as $\sqrt{\lambda}$ and as the third power of the temperature,
   \be
\hat q_{\rm conformal}={\Gamma\left[{3\over 4}\right]\over \Gamma\left[{5\over 4}\right]}~\sqrt{2\l}~\pi^{3\over 2}T^3
\ee

A different approach chooses to compute the drag force acting a string whose UV end-point
 (representing an infinitely heavy quark) is forced to move with constant velocity $v$, \cite{her,gub1,tea}, in the context of
 ${\cal N}=4$ SYM plasma.
 The result for the drag force is
 \be
F_{\rm conformal}={\pi\over 2}\sqrt{\lambda}~T^2{v\over \sqrt{1-v^2}}\ee
and is calculated by first studying the equilibrium configuration of the appropriate string world-sheet string
and then calculating the momentum flowing down the string.
This can be the starting point of a Langevin evolution system,  as the process of
energy loss has a stochastic character, as was first pointed out in \cite{lan1} and more recently pursued in \cite{lan2}-\cite{iancu}.

Such a system involves a classical force, that in this case is the drag force, and a stochastic noise that is taken to be Gaussian
and which is characterized by a diffusion coefficient.
There are two ingredients here that are novel. The first  is that the Langevin evolution must be relativistic, as
the quarks can be very energetic. Such relativistic systems have been described in the mathematical physics literature, \cite{math}
and have been used in phenomenological analyses of heavy-ion data, \cite{lan}.
They are known however to have peculiar behavior, since demanding an equilibrium relativistic Boltzmann
 distribution, provides an Einstein relation that is pathological
at large temperatures. Second, the transverse and longitudinal
diffusion coefficients are not the same, \cite{Gubser-lan}. A
first derivation of such Langevin dynamics from holography was
given in \cite{Gubser-lan}. This has been extended in in
\cite{iancu} where the thermal-like noise  was associated and interpreted in
terms of the world-sheet horizon that develops on the probe
string.

Most of the transport properties mentioned above have been successfully computed in ${\cal N}=4$ SYM and a lot of
debate is still waged as to how they can be applied to QCD in the appropriate temperature range, \cite{gub2},\cite{caron},\cite{sin}.
A holographic description of QCD has been elusive,  and the best we have so far have been simple bottom up models.

In the simplest bottom-up holographic model known as  AdS/QCD \cite{adsqcd1},
the bulk viscosity is zero as conformal invariance is essentially not broken
(the stress tensor is traceless),  and the drag force and jet quenching essentially retain their conformal values.

In the soft-wall model \cite{soft}, no reliable calculation can be done for glue correlators and therefore
transport coefficients are ill-defined, as bulk equations of motion are not respected. Similar
remarks hold for other phenomenologically interesting observables as the
drag force and the jet quenching parameter.

A hybrid approach has been advocated in \cite{ihqcd1,ihqcd2,k}
combining features of bottom-up and top-down (string theory)
models. Such an approach is essentially a five-dimensional
dilaton-gravity system with a non-trivial dilaton potential.
Flavor can be eventually added in the form of $N_f$ space-time
filling $D4-\overline{D4}$ brane pairs, supporting $U(N_f)_L\times
U(N_f)_R$ gauge fields and a bi-fundamental scalar
\cite{ckp}\footnote{$D4-\overline{D4}$ brane pairs for flavor
where first suggested in \cite{bi} and the finite temperature
solutions studied in \cite{edel}.}.

The UV asymptotics of the potential are fixed by QCD perturbation theory, while the
 IR asymptotics of the potential can be fixed by confinement and linear
 glueball asymptotics.

An analysis of the finite temperature behavior \cite{gkmn1,gkmn2} has
shown that the phase structure is exactly
 what one would expect from large-$N_c$ YM\footnote{Similar results, but with
 somewhat different potentials were also obtained in \cite{nellore,dew}.}.
Einstein-dilaton gravity with a strictly monotonic dilaton potential that grows sufficiently
fast, generically shares the same phase structure and
thermodynamics of finite-temperature pure Yang-Mills theory at large $N_c$.
There is a deconfinement phase transition (dual to a Hawking-Page phase transition between a black-hole and thermal gas
background on the gravity side), which is generically first order.
 The latent heat scales as $N_c^2$. In the deconfined gluon-plasma phase,
the free energy slowly approaches that of a free gluon gas at high temperature,
and the speed of sound starts from a small value at $T_c$ and approaches the conformal value $c_s^2=1/3$
 as the temperature increases. The deviation from conformal invariance is strongest at $T_c$, and
is signaled by the presence of a non-trivial gluon condensate, which on the gravity side emerges as
a deviation of the scalar solution that behaves asymptotically as $r^4$ close to the UV boundary. In the CP-violating
sector, the topological vacuum density $ tr~F\tilde{F}$ has zero expectation value in the deconfined phase,
in agreement with lattice results \cite{ltheta} and large-$N_c$ expectations.

The analysis performed in \cite{gkmn2} was completely general and did not rely on any specific form of the dilaton potential $V(\l)$.
A potential with two parameters, was subsequently chosen to describe YM data, \cite{gkmn3}.
The (dimensionless) free energy, entropy density, latent heat and speed of sound, obtained on the gravity side by numerical integration of the 5D
field equations, were compared with the corresponding quantities, calculated on the lattice  for pure Yang-Mills at finite-$T$,
resulting in excellent
agreement, for the temperature range that is accessible by lattice techniques.
The same  model  also shows a good agreement with the lattice calculation of glueball mass ratios at zero temperature.
Moreover the  value of the deconfining
critical temperature (in units of the lowest glueball mass) was  also in good agreement with the lattice results.

In short, the model, named Improved Holographic QCD, (or IHQCD for short),  gives a good phenomenological (holographic) description of
most static  properties (spectrum and equilibrium thermodynamics)  of
large-$N_c$ pure Yang-Mills, as computed on the lattice, for energies up to several  times $T_c$.
 Therefore it constitutes a good starting point for the
computation of dynamical observables in a {\em realistic} holographic dual to QCD (as opposed to e.g. ${\cal N}=4$ SYM), such as transport
coefficients and other hydrodynamic properties that are not easily accessible by lattice techniques, at energies and temperatures relevant for
relativistic heavy-ion collision experiments.

The purpose of the present paper is to compute transport properties
(the bulk viscosity) and energy loss coefficients (the jet quenching parameter and the drag
force) in the specific  Improved Holographic QCD model  described in \cite{gkmn3}.

The shear viscosity of IHQCD is the same as that of ${\cal N}=4$
SYM, as the model is a two derivative model. Although this is not
a good approximation in the UV of QCD, it is expected to be a good
approximation in the energy range $T_c\leq T\leq 5T_c$. We find
that the bulk viscosity rises near the phase transition
 but ultimately stays slightly below the shear viscosity. We also
 give a general holographic argument that any (large-N) gauge theory that confines color at zero temperature should
 have an increase in the bulk viscosity-to-entropy density ratio close to
 $T_c$.

The drag force on heavy quarks,  and the associated diffusion times, are calculated and found to be momentum depended as anticipated from asymptotic freedom.
Numerical values of diffusion times are in the region dictated by phenomenological analysis of heavy-ion data.
We calculated the medium-induced corrections to the  quark mass (needed
for the diffusion time calculation),
and we find they result in a mildly decreasing  effective quark  mass
as a function of temperature. This is consistent with lattice results.
Finally the jet-quenching parameter is calculated and found to be comparable at $T_c$ to the one obtained by extrapolation from
${\cal N}=4$ SYM. Its temperature dependence is however different and again reflects the effects of asymptotic freedom.

There are several sources of error and systematic uncertainties in the results above.
We analyze them in the appropriate sections and make a long commentary on their importance
in the last section.

The structure of the paper is as follows. In Section 2, we
review the holographic construction that shall be used to compute
certain observables of QGP. In particular we review the asymptotic
behaviors of the backgrounds and discuss the various parameters in
the model and how they are fixed. Section 3 is devoted to the
computation of the bulk viscosity. We describe the general
holographic computation of the quantity from the graviton
fluctuation equations on the dual background and compute its
temperature dependence numerically. We also make a proposal for a
holographic explanation of the rise in the bulk viscosity near the
phase transition. In section 4, we compute the drag force on a
heavy quark moving in QGP in our set-up. We obtain general
analytic formulas in the relativistic and the non-relativistic
limits of the drag force as a function of temperature. We compare
our findings with the ${\cal N}=4$ SYM result. In this section we
also compute the diffusion times for the heavy quarks in the QGP.
In particular, we present numerical results for the charm and the
bottom quarks. Furthermore, we compute the thermal corrections to
the quark masses in our set-up and finally discuss in what
temperature ranges should our results be trusted. In section 5, we
compute the jet-quenching parameter in our set-up. Our findings
are compared with the conformal (${\cal N}=4$ SYM) case. Finally,
the section 6 contains a discussion and outlook. The various
appendices detail our computations.

\section{Review of IHQCD backgrounds}

The holographic duals of large $N_c$ Yang Mills theory proposed in \cite{ihqcd1,ihqcd2} are
based on five-dimensional Einstein-dilaton gravity with a dilaton potential. The basic fields
for the pure gauge sector  are the
5D metric $g_{\mu\nu}$ (dual to the 4D stress tensor) and a scalar field $\Phi$ (dual
to $Tr F^2$). The Einstein frame action for these fields is:
\begin{equation}
   {\cal S}_5=-M^3_pN_c^2\int d^5x\sqrt{g}
\left[R-{4\over 3}(\partial\Phi)^2+V(\Phi) \right]+2M^3_pN_c^2\int_{\partial M}d^4x \sqrt{h}~K.
 \label{a1}\end{equation}
Here, $M_p$ is the  five-dimensional Planck scale and $N_c$ is the
number of colors. The last term is the Gibbons-Hawking term, with
$K$ being the extrinsic curvature on the boundary. The effective
five-dimensional Newton constant is $G_5 = 1/(16\pi M_p^3 N_c^2)$,
and it is small in the large-$N_c$ limit.

The scalar potential $V(\Phi)$ is what determines the dynamics. Its form  is in part motivated
from non-critical string theory, and in part chosen following guidelines from phenomenology.
We will often write $V$ as a function of $\l \equiv e^\Phi$.

Asymptotic freedom in the UV requires $V(\l)$  have a regular
expansion for small $\l \equiv e^\Phi$: \be \label{VUV} V(\l) =
{12\over \ell^2} \left(1 + V_0 \l + V_1 \l^2 + \ldots\right),
\quad V_0>0,  \qquad \l \to 0. \ee This ensures that in any
solution of Einstein's equations the metric has an asymptotically
$AdS_5$  UV region, with $AdS$ length $\ell$, in which the field
$\l$ vanishes logarithmically. We have the perturbative
identification, valid for small $\l$: \be \kappa \l \equiv  N_c
g^2_{YM}, \qquad \kappa \equiv {9 \over 8} {V_0 \over \beta_0},
\ee where $\beta_0 = (22/3)(4\pi)^{-2}$ is the first coefficient
of the perturbative beta-function for the 't Hooft coupling $N_c
g_{YM}^2$ of  pure Yang-Mills theory.

For large $\l$, confinement and a linear glueball spectrum
require: \be\label{VIR} V(\l) \simeq V_{\infty} \l^{4/3} (\log
\l)^{1/2} \qquad \l \to \infty, \ee where $V_{\infty}$ is a
positive constant\footnote{Other types of large-$\l$ asymptotics
also lead to color confinement, with different features of the
glueball spectrum. These solutions were analyzed in complete
generality in \cite{ihqcd2}.  }.

For a generic potential that obeys the asymptotics (\ref{VUV}) and (\ref{VIR}),
the model exhibits the following features:
\begin{itemize}

\item  \paragraph{Zero temperature.}
At zero temperature, the gravity solution  is dual to a confining
4D theory. One has  color confinement (i.e. a Wilson Loop area
law) and a discrete glueball spectrum with linear behavior, $m_n^2
\sim n$. The solution of Einstein's equations for the  metric and
dilaton has the form: \be\label{confsol} ds^2 = e^{2A_o(r)}
\left(dr^2 + \eta_{\mu\nu} dx^\mu dx^\nu\right) , \qquad \l =
\l_o(r), \qquad 0< r<\infty, \ee with small-$r$ log-corrected
$AdS$ asymptotics, \be\lab{logAdS} A_o(r) \sim -\log r/\ell +
{\cO}\le(\frac{1}{\log \Lambda r}\ri) + \ldots, \qquad \l_o(r)
\sim -\frac{1}{\log \Lambda r}
\ee
 and large-$r$ behavior:
\be \label{metricIR} A_o(r) \sim - C r^2, \qquad \l_o(r) \sim
\exp\left[{3\over 2} C r^2\right], \qquad r\to \infty.  \ee In equation (\ref{logAdS}),
$\Lambda$ is an integration constant that sets the length scale of
nonperturbative physics;  the constant  $C$ in equation (\ref{metricIR})  is determined in terms of $\Lambda$.

\item \paragraph{Finite  temperature.} At finite temperature,  one finds a first order phase transition between a low-temperature
 confined phase, described by the  solution (\ref{confsol}), and a high-temperature
deconfined phase, described holographically by a 5D black hole
solution:
\be
ds^2=e^{2A(r)}\left[{dr^2\over f(r)}-f(r)dt^2+dx_mdx^m\right], \qquad \l = \l(r) , \qquad 0< r< r_h.
 \label{BH1}\ee
These solutions are characterized by the presence of a horizon $r_h$ where $f(r_h)=0$, and
have a  temperature $T$ and an  entropy density $s$:
\be\label{TS}
 T = - {\dot{f}(r_h) \over 4\pi}, \qquad s  = 4\pi\, (M_p^3 N_c^2) \, e^{3A(r_h)}.
\ee
In the UV ($r\to 0$), and for any $r_h$ , the black holes are asymptotically $AdS_5$ and
reduce to the zero-temperature metric $A(r)\simeq A_o(r)$,  $f(r)\simeq 1$.

\end{itemize}

In all types of solutions,  (\ref{confsol}) and (\ref{BH1}),  the dilaton $\Phi(r)$ is a monotonically increasing function of $r$. One can therefore  use
$\Phi$ itself as the radial coordinate in (\ref{BH1}):
\be ds^2 =e^{2A(\f)}(-f dt^2 +dx_mdx^m) + e^{2B(\f)}\frac{d\f^2}{f}.
 \label{BH2}\ee
 Comparison of (\ref{BH1}) and (\ref{BH2}) determines,
\be\label{B11} B = A -\log\Big|\frac{d\f}{dr}\Big|.\ee This form of the
metric will prove useful later.

Generically, in these types of models
there exist two separate black-hole solutions, that were referred
in \cite{gkmn2} as the {\em big} and the {\em small} black-holes.
In \cite{gkmn2} it was proved that existence of this
second type of black-hole solution (the small BH) is necessary and
sufficient for a first order confinement-deconfinement phase
transition.

The big BH solution exists for $T>T_{min}$ for some finite
$T_{min}$,
 see figure \ref{illus} (b), and dominates the
entire thermodynamic ensemble for $T>T_c$ where $T_c$ is always
larger than $T_{min}$. It always dominates in the thermodynamic
ensemble over the small BH. It corresponds to the range
$0<\l_h<\l_{min}$ in the horizon value of the dilaton, for some
finite $\l_{min}$, see figure \ref{illus} (b). This solution is
proposed as the holographic dual of the Yang-Mills gluon plasma.

The small BH solution also exists for $T>T_{min}$ and it
corresponds to the range $\l_{min}<\l_h<\infty$, see figure
\ref{illus} (b). As it is never dominant in the thermodynamic
ensemble, it bears no direct significance for an holographic
investigation of the quark-gluon plasma.

\begin{figure}[ht]
\centering
\includegraphics[scale=1]{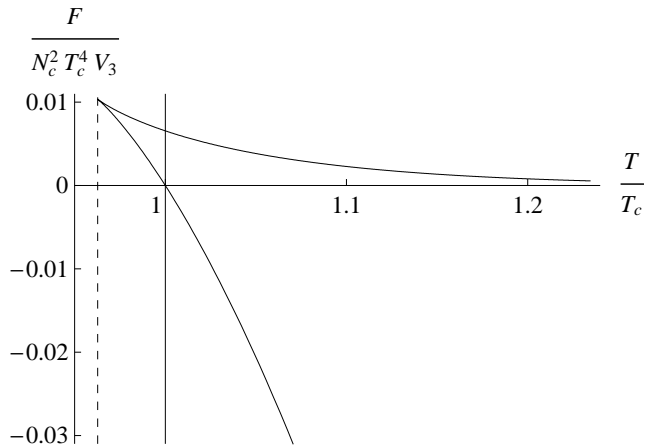}
\caption{Free energy of black hole solutions as a function of temperature. The
(constant) free energy of the $T=0$ confining  vacuum is set to zero.
The two branches correspond to the big black holes (lower branch) and the small black holes (upper branch).  The two branches merge at a minimum
temperature $T_{min} >0$, corresponding to the vertical dashed line.
The free energy of the big black hole  branch crosses the x-axis at $T=T_c$, indicating a first order phase transition between the vacuum and big black hole phase.}\label{FT}
\end{figure}

This situation is depicted in Figure \ref{FT}.

In summary, there exists three separate solutions to the
dilaton-gravity system:
\begin{itemize}
\item[i.] The thermal gas (\ref{confsol}) that exists for all $T>0$. It
is the dominant solution for $T<T_c$.
\item[ii.] The big BH (\ref{BH1}) that exists for $T>T_{min}$ and
 becomes the dominant solution for $T>T_c$.
\item[iii.] The small BH that exists for $T>T_{min}$ and is always
sub-leading in the thermodynamic ensemble.
\end{itemize}
%This phase structure  is shown in Figure \ref{FT}.

The solutions (\ref{confsol}) and (\ref{BH1}) are written in the Einstein frame. Some
of the transport properties we compute
 in this paper however are defined in terms
of the string frame,
 since they are  related to world-sheet quantities.
In the five-dimensional non-critical string setup,
the string frame and the Einstein frame metrics are related by \cite{ihqcd1}:
\be\label{stfr}
ds^2_s = e^{{4\over 3}\Phi} ds^2_E,
\ee
so we can define a string frame scale factor (both at zero and finite temperature):
\be\label{stfr2}
A_s(r) = A(r)  + {2\over 3} \Phi(r).
\ee

As shown in \cite{ihqcd2}, the Einstein frame scale factor is  monotonic if
the metric is asymptotically $AdS$ and  the theory satisfies the null energy condition\footnote{This is always the case for a single scalar field with a canonical kinetic term.}. On the other hand,
the string frame scale factor may not be monotonic. In particular,
 in the  backgrounds with IR asymptotics (\ref{metricIR}), (which follows
 if the dilaton potential obeys (\ref{VIR}) ) the zero-temperature string frame
scale factor behaves as:
\be\label{stfr3}
A_{s,o}(r) \sim \left\{\begin{array}{l} -\log r/\ell \to +\infty \qquad  r\to 0, \\
\;\;{3\over 4}\log r \to +\infty \qquad r\to \infty.
\end{array}\right.
\ee
Therefore, the zero-temperature string frame  scale scale factor must have
a minimum at some finite
value of the radial coordinate, $r = r_*$, where in string units the metric has a minimum size
$e^{A_{s,o}(r_*)}$.     This is what causes the holographic Wilson loop
to exhibit an area law \cite{cobi}. The confining string tension $\sigma_c$ is related to
the fundamental string length  $\ell_s$ and  value of the string frame metric at the extremum:
\be\label{strten}
\sigma_c =  {e^{2A_{s,o}(r_*)}\over 2\pi \ell_s^2}       .
\ee

Notice that it is not guaranteed that the minimum survives in the
black hole solutions. For sufficiently high temperature, the
minimum of the string world-sheet should disappear behind the
horizon. In fact, this is what happens in the explicit case we
will consider in this paper: a numerical analysis shows that for
all temperatures larger than the critical temperature $T_c$, both
the string and Einstein frame scale  factor are monotonically
decreasing  over the whole range $0<r<r_h$. Thus the minimum of
the scale factor in both frames occurs at $r=r_h$.

In \cite{gkmn3} we assumed a specific form of the potential:
\be\lab{dilpot} V(\l)  = {12\over \ell^2} \left\{ 1 + V_0 \l + V_1
\l^{4/3} \left[\log \left(1 + V_2 \l^{4/3} + V_3 \l^2\right)
\right]^{1/2} \right\}. \ee

The model specified by the potential (\ref{dilpot}) contains a  few adjustable parameters, namely the
coefficients $V_i$ and $\ell$ entering the potential, and the 5D Planck scale $M_p$.
They  were fixed in \cite{gkmn3} as follows:

\begin{itemize}

\item The coefficients $V_0$ and $V_2$  are chosen to reproduce the
perturbative Yang-Mills beta-function up to 2-loop order, $\beta(\l) =
-\beta_0 \l^2 - \beta_1 \l^3 + O(\l^4)$.
  This requires:
\be
V_0 = {8\over 9} \beta_0, \qquad V_2 = \beta_0^4 \left({23  + 36\,
\beta_1/\beta_0^2 \over 81 V_1 }\right)^2.
\ee
For pure Yang-Mills the beta-function coefficients are:
\be
\beta_0 = {22\over 3 (4 \pi)^2}, \qquad \beta_1 = {51\over 121} \beta_0^2.
\ee

\item
The coefficients $V_1$ and $V_3$ were fixed by comparing the
latent heat of the phase transition,
and the pressure  of the deconfined phase at a given  temperature ($T=2T_c$), to the corresponding lattice results.  A successful matching leads to the choice:
\be
V_1 = 14, \qquad V_3 = 170.
\ee

\item
The asymptotic $AdS$ scale $\ell$ only affects the overall unit of
energy, and can be set by fixing the value of a single
dimensionful quantity in the model (say the lowest glueball mass,
or the critical temperature). Any physical dimensionless quantity
is independent of $\ell$. Once $\ell$ is given, the UV solution is
asymptotically: \be\label{solUV}
  A(r) = \log {\ell\over r} + O\left({1\over \log r}\right), \qquad
\l(r) = -{1\over \b_0 \log \Lambda r} +  O\left({\log \log r\over
\log^{2}r}\right). \ee The scale $\Lambda$  appearing in the UV
asymptotics of $\lambda(r)$ is an integration  constant  of the
zero-temperature Einstein's equations, and it is  related to the
UV boundary conditions $(A_0, \l_0)$  at a small but finite
coordinate $r_0$ as: \be \Lambda \simeq \ell^{-1}(\b_0 \l_0)^{-b}
\exp\left\{A_0 - {1\over \b_0 \l_0}\right\}, \qquad b = {\b_1\over
\b_0^2}. \ee It may seem from this discussion that there is an
extra dimensionless parameter in our model, $\Lambda \ell$, with
respect  to pure 4D Yang-Mills (where the only parameter is the
scale $\Lambda$). This is not so: all physical quantities that can
be related holographically to a Yang-Mills observable have a
trivial dependence on $\Lambda \ell$. In fact, as shown in
\cite{ihqcd2},  changing $\Lambda$ while keeping  $\ell$ fixed,
is the same as shifting $A(r)$ by a constant, i.e.  a fixed
rescaling of all energies in the model or a change of units.
  Alternatively, for any given value of $\ell$, there
exists a unique solution such that
the scale $\Lambda$ is equal to the physical value in 4D Yang-Mills,
and that  no dimensionless observable
depends on this choice.

\item The 5D Planck scale is fixed (in units of $\ell$) so that, in
the $T\to \infty$ limit,
the equation of state matches that of a free relativistic gas of
$N_c^2$ photons,
\be
\lim_{T\to \infty}{p(T) \over T^4}= {\pi^2 \over 45} \quad
\Leftrightarrow \quad M_p^3 = {\ell^{-3} \over 45 \pi^2}.
\ee

\end{itemize}

As shown in \cite{gkmn3}, with these choices of the parameters the 5D holographic model
is able to accurately reproduce all known thermodynamic properties of  finite temperature
Yang-Mills theory, as they emerge from lattice studies. It also displays a glueball spectrum
which is in good agreement with lattice results. The  value of the
confinement-deconfinement transition is found  to be $T_c = 247$ MeV,
very close to the lattice determination of the YM critical temperature.

In the following sections  we discuss  the transport coefficients (i.e. the
bulk viscosity) of the deconfined phase  and the  energy loss of a heavy quark   in this specific  holographic model.

\section{Bulk viscosity}\label{BULKVISC}

The bulk viscosity $\zeta$ is  an important probe of the quark-gluon plasma.
Its profile as a function of T reveals information
regarding the dynamics of the phase transition. In particular, both from the
low energy theorems and lattice studies \cite{Viscos1,Viscos2,Viscos3}, there is evidence
that $\zeta$ increases  near $T_c$.

For a viscous fluid the shear $\eta$ and bulk $\zeta$ viscosities are defined via the rate of entropy production as
\be
{\partial s\over\partial t}={\eta\over T}\left[\partial_iv_j+\partial_jv_i-{2\over 3}(\partial\cdot v)\delta_{ij}\right]^2
+{\zeta\over T}(\partial\cdot v)^2
\ee

Therefore, in a holographic setup, the bulk viscosity can be defined
as the response of the diagonal spatial components of the
stress-energy tensor to a small fluctuation of the metric.
It can be directly related to the retarded Green's
function of the stress-energy tensor by Kubo's linear response
theory:
\be \label{zeta1} \z=-\frac19 \lim_{\omega \rightarrow 0}
\frac{1}{\omega} Im G_R(\omega,0),
\ee
 where $G_R(w,\vec{p})$ is the
Fourier transform of  retarded Green's function of the
stress-energy tensor:
\be\lab{green} G_R(w,\vec{p}) = -i\int
d^{3}x dt e^{i \omega t - i \vec{p}\cdot\vec{x}}\theta(t)
\sum^{3}_{i,j=1}\langle\,[T_{ii}(t,\vec{x}),T_{jj}(0,0)]\,\rangle. \ee
A direct
computation of the RHS on the lattice is non-trivial as it
requires analytic continuation to Lorentzian space-time. In refs.
\cite{Viscos1},\cite{Viscos2}  the low energy theorems of QCD, as well as (equilibrium) lattice data at
finite temperature were used in order to evaluate a particular moment of the
spectral density of the relevant correlator.
using a parametrization of the spectral density via two time-dependent constants, one of which is the bulk viscosity
a relation for their product was obtained as a function of temperature.
This can be converted to a relation for $\zeta$, assuming the other constant varies slowly with temperature.

The conclusion was that  $\zeta/s$
increases near $T_c$. Another conclusion is that the fermionic contributions to $\zeta$ are small compared to the glue contributions.

The  weak point of the approach of
\cite{Viscos2}, is that  it requires an ansatz on the spectrum of
energy fluctuations, and further assumptions on the other parameters.
which are not derived from first principles.

A direct lattice study of the bulk viscosity was also made in
\cite{Viscos3}. Here, the result is also qualitatively similar
\ref{ViscosFigI}. However, the systematic errors in this computation are large
especially near $T_c$,  mostly due to the analytic continuation that
one has to perform after computing the Euclidean correlator on the
lattice.

The results of references \cite{Viscos1},\cite{Viscos2} and the
assumptions of the lattice calculation have been recently challenged in
\cite{ms}.

\subsection{Holographic computation and main results}

The holographic approach offers a new way of computing the bulk viscosity.
 In the
holographic set-up, $\zeta$ is obtained from (\ref{zeta1}). Using
the standard AdS/CFT prescription, the two point-function of the
energy-momentum tensor can be read off from the asymptotic
behavior of the metric perturbations $\delta g_{\mu \nu}$. This is
similar in spirit to the holographic computation of the shear
viscosity \cite{Shear}, but it is technically more involved. For a
recent treatment of the fluctuation equation governing the scalar
mode  of a general  Einstein-Dilaton system, see \cite{springer}.
Here, we shall follow the method proposed by \cite{Gubser:2008sz}.

As explained in \cite{Gubser:2008sz}, one only needs to examine
the equations of motion in the gauge $r=\phi$, where the radial
coordinate is equal to the dilaton. In our type of metrics, the applicability of this
method requires some clarifications, that we provide  in App. \ref{GaugeEquivalence}.
Using $SO(3)$ invariance
and the five remaining gauge degrees of freedom the metric
perturbations can be diagonalized as \be\label{pert1} \delta g =
diag(g_{00},g_{11},g_{11},g_{11},g_{55}), \ee where
\be\label{pert2} g_{00} = -e^{2A}f [1+ h_{00}(\f)e^{-i\o t}],\quad
g_{11} = e^{2A} [1+ h_{11}(\f)e^{-i\o t}],\quad g_{55} =
\frac{e^{2B}}{f}[1+ h_{55}(\f)e^{-i\o t}],\ee where the functions $A$ and $B$ are defined in (\ref{BH2}).
Here, the fluctuations are taken to be harmonic functions of $t$ while having an
arbitrary dependence on $\f$.

The bulk viscosity depends only on the correlator of the diagonal
components of the metric and so it suffices to look for the
asymptotics of $h_{11}$.  Interestingly, in the $r=\phi$ gauge
this decouples from the other components of the metric and
satisfies the following equation\footnote{Difference in the
various numerical factors in this equation w.r.t
\cite{Gubser:2008sz} is due to our different normalization of the
dilaton kinetic term.} \be \label{hw} h_{11}''
-\left(-\frac{8}{9A'}-4A'+3B'-\frac{f'}{f}\right)h_{11}'
-\left(-\frac{e^{2B-2A}}{f^2}\omega^2
+\frac{4f'}{9fA'}-\frac{f'B'}{f}\right)h_{11} = 0\;. \ee One needs
to impose two boundary conditions. First, we require that only the
infalling condition survives at the horizon: \be \label{infall}
h_{11} \to c_b (\f_h-\f)^{-\frac{i\omega}{4\pi T}}, \qquad
\f\to\f_h, \ee where $c_b$ is a normalization factor. The second
boundary condition is that $h_{11}$ has unit normalization on the
boundary: \be\label{unitnorm} h_{11}\to 1, \qquad \f\to-\infty.
\ee Having solved for $h_{11}(\f)$, Kubo's formula (\ref{zeta1})
and a wise use of the AdS/CFT prescription to compute the
stress-energy correlation function \cite{Gubser:2008sz} determines
the ratio of bulk viscosity as follows.

The AdS/CFT prescription relates the imaginary part of the retarded
$T_{ii}$ Green function to the number flux of the $h_{11}$
gravitons ${\cal F}$ \cite{Gubser:2008sz}:
\be\lab{green2} Im\,
G_R(\omega,0) = -\frac{\cal F}{16\pi G_5} \ee where the flux can be
calculated as the Noether current associated to the $U(1)$
symmetry $h_{11}\to e^{i\theta} h_{11}$ in the gravitational
action for fluctuations. One finds, \be\lab{flux1} {\cal F}  =
i\frac{e^{4A-B}f}{3A^{'2}}[h_{11}^*h_{11}'-h_{11}h_{11}^{*'}]. \ee
As ${\cal F}$ is independent of the radial variable, one can compute it at
any $\f$, most easily near the horizon, where $h_{11}$ takes the
form (\ref{infall}). Using also the fact that $(dA/d\f )(\f_h) =
-8V(\f_h)/9V'(\f_h)$ (see Appendix \ref{AppA}), one finds
\be
\lab{flux2} {\cal F(\o)} = \frac{27}{32} \o |c_b(\o)|^2
e^{3A(\f_h)}\frac{V'(\f_h)^2}{V(\f_h)}.
\ee
Then, (\ref{zeta1})
and (\ref{green2}) determine the ratio of bulk viscosity and the
entropy density as,
\be
\label{zs} \frac{\zeta}{s} =
\frac{3}{32\pi} \left(\frac{V'(\f_h)}{V(\f_h)}\right)^2 |c_b|^2.
\ee
In the derivation we use the Bekenstein-Hawking formula for
the entropy density, $s = \exp{3A(\f_h)}/4G_5$.

To find $\zeta$ we need to find $c_b$ only in the limit $\omega\to 0$.
The computation is performed by numerically solving equation (\ref{hw})
with the appropriate boundary conditions.
There are two separate methods that one can employ to
determine the quantity $c_b$:
\begin{enumerate}
\item
One can solve (\ref{hw}) numerically with a fixed $\omega/T$, but small enough so
that $c_b$ reaches a fixed value. The method is valid also for
finite values of $\omega$. From a practical point of view, it is easier
to solve (\ref{hw}) with the boundary condition (\ref{infall})
with a unit normalization factor, read off the value on the
boundary $h_{11}(-\infty)$ from the solution and finally use the
symmetry of (\ref{hw}) under constant scalings of $h_{11}$ to
determine $|c_b| = 1/|h_{11}(-\infty)|$.

\item An alternative method of computation that directly extracts
the information at $\omega=0$ follows from the following trick
\cite{Gubser:2008sz}. Instead of solving (\ref{hw}) for small but
finite $\omega$, one can instead solve it for $\omega=0$. This is a simpler
equation, yet complicated enough to still evade analytic solution. Let
us call this solution $h_{11}^0$.
 One numerically solves it by fixing the boundary conditions on the boundary:
$h^0_{11}(-\infty)=1$ and the derivative $dh^0_{11}/d\f (-\infty)$ is chosen such that $h_{11}$ is regular at the horizon.
Matching this solution to the expansion of (\ref{infall}) for small $\omega$
than yields $|c_b| = h^0_{11}(\f_h)$.
\end{enumerate}

\begin{figure}
 \begin{center}
\includegraphics[height=10cm,width=12cm]{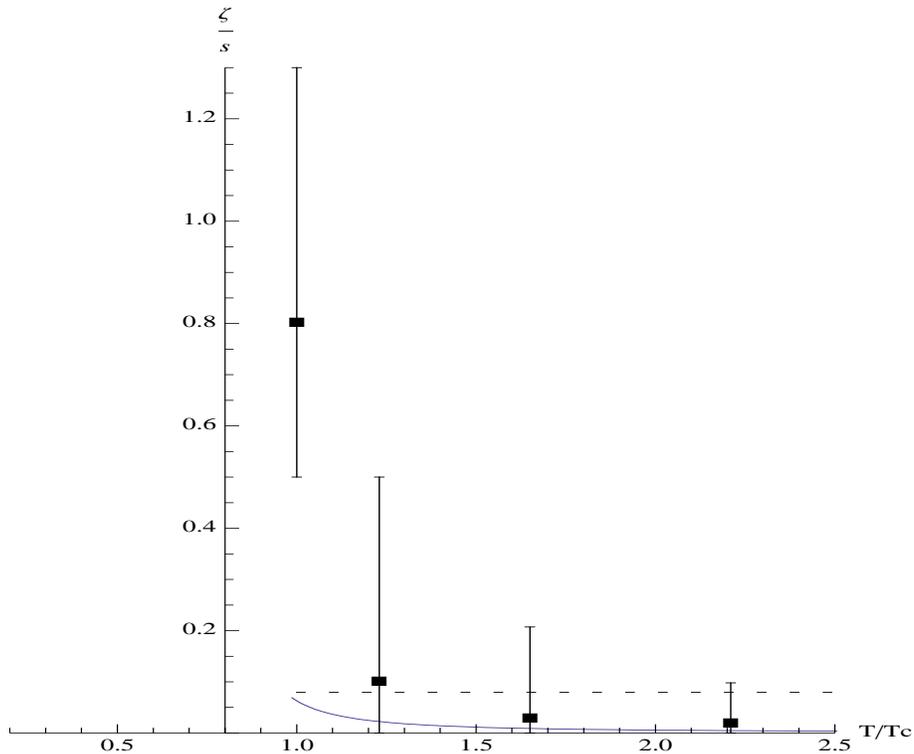}
 \end{center}
 \caption[]{Plot of $\zeta/s$ (continuous line) calculated in Improved Holographic QCD model.
  This is compared with the lattice data of \cite{Viscos3} that are shown as boxes. The horizontal dashed line is
  indicating the (universal) value of ${\eta\over s}$ for comparison. }
\label{ViscosFigI}
 \end{figure}

We used both methods to obtain $\zeta/s$ as a function of T and
checked that they yield the same result. As explained in
\cite{gkmn2}, most of the thermodynamic observables are easily
computed using the method of scalar variables. This method is
summarized in appendix {\ref{AppA} where we also detail the computation of
the bulk viscosity using these variables.

Our results are
presented in figure \ref{ViscosFigI}. This figure gives a comparison
of the curve  obtained by the holographic calculation sketched above
by  solving (\ref{hw}) and the lattice data of \cite{Viscos3}. We also
show $\eta/s=1/4\pi$ in this figure for comparison. The result is
qualitatively similar to the lattice result where $\zeta/s$
increases as $T$ approaches $T_c$, however the rate of increase is
slower than the lattice. As a result, we obtain a value
$\zeta/s(T_c)\approx 0.06$ that is an order of magnitude smaller
than the lattice result\cite{Viscos3} which is 0.8. Note however
that the error bars in the lattice evaluation are large near $T_c$
and do not include all possible systematic errors from the
analytic continuation.

We should note the fact that the holographic calculation gives a
 smaller value for the bulk viscosity near $T_c$
than the lattice calculation is generic and has been found for
other potentials with similar IR asymptotics,
\cite{Gubser:2008sz}. The fact that the value of $\z/s$ near $T_c$ is
 correlated with the IR asymptotics of the potential  will
be shown further below.

Another fact that one observes from figure \ref{ViscosFigI} is
that $\zeta/s$ vanishes in the high $T$ limit. This reflects the
conformal invariance in the UV and can be shown analytically as
follows. $\z/s$ is determined by formula (\ref{zs}). In the high T
limit, (corresponding to $\l_h\to 0$, near the boundary), the
fluctuation coefficient $|c_b|\to 1$. This is because of the
boundary condition $h_{11}(\l=0)=1$. We use the relation between T
and $\l_h$ in the high T limit \cite{gkmn2}, \be \lab{lhThighT}
\l_h\to \le(\beta_0\log(\pi T/\Lambda)\ri)^{-1}. \ee Substitution
in (\ref{zs}) leads to the result, \be \lab{highTbBH}
\frac{\z}{s}\bigg|_{big}\to \frac{2}{27\pi}\frac{1}{\log^2(\pi
T/\Lambda)}, \qquad as\,\,\, T\to\infty. \ee As $s$ itself
diverges as $T^3$ in this limit -- it corresponds to an ideal gas
--  we learn  that $\zeta$ also diverges as $T^3/\log^2(T)$.
Divergence at high T is expected from the bulk-viscosity of an
{\em ideal gas}. We do not expect however the details of the
asymptotic result to match  with the pQCD result, for the same
reasons that the shear-viscosity-to-entropy ratio does not,
\cite{k}. However, the T-dependence is very similar to the pQCD
result,  \cite{Arnold}: \be \zeta/s \propto \log^{-2}(\pi
T/\Lambda)\log^{-1}\log(\pi T/\Lambda)\;. \ee

\subsection{Holographic explanation for the rise of $\zeta/s$ near $T_c$ and the small black-hole branch}

 With the same numerical methods, one can also compute the ratio $\zeta/s$ on the small black-hole
 branch. As this solution has a smaller value of the
 action than the big black-hole solution, it is a subleading saddle point in the phase space of the theory,
 hence bears no direct significance for an holographic investigation of the
 quark-gluon plasma. However, as we show below, the existence of this branch
 provides  a holographic explanation for the peak
 in $\zeta/s$ in the quark-gluon plasma, near $T_c$.

\begin{figure}
 \begin{center}
 \leavevmode \epsfxsize=7cm \epsffile{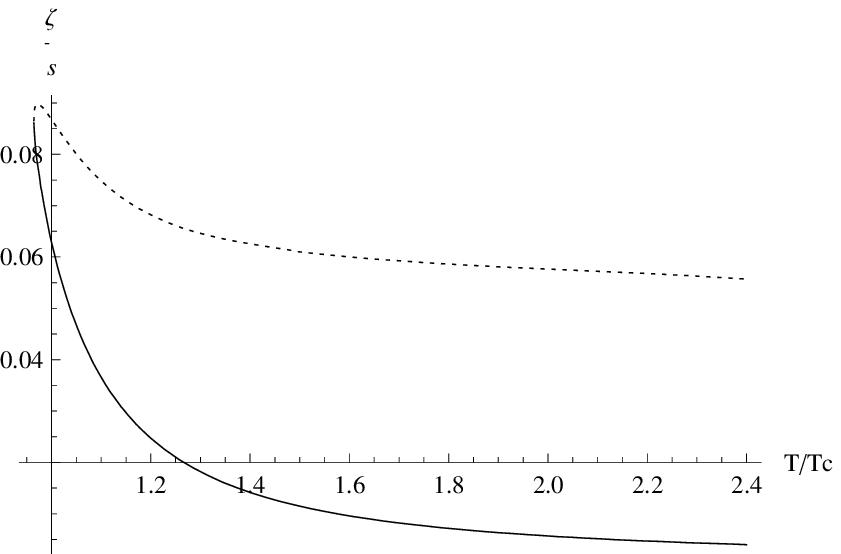}
  \leavevmode \epsfxsize=7cm \epsffile{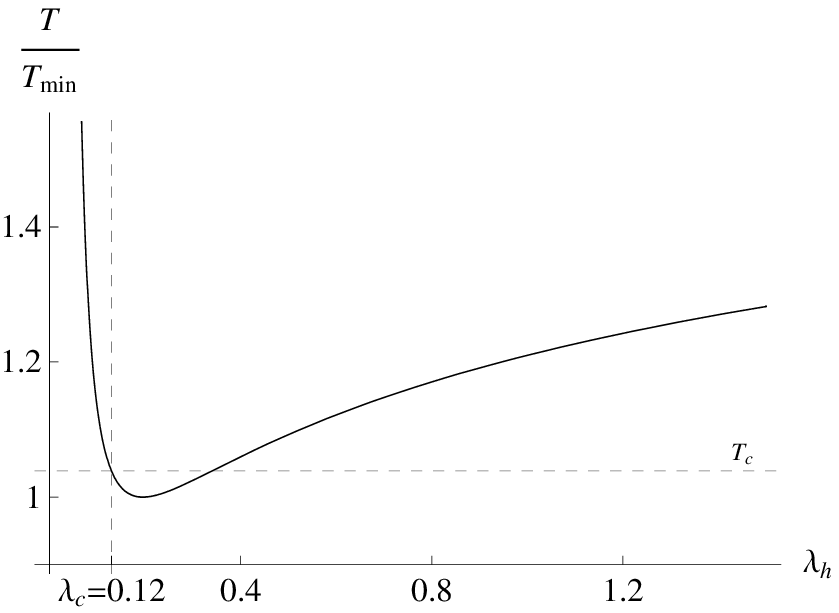}
 \end{center}
 \vspace*{3pt}
 \centerline{\hspace{.3in} {\bf (a)} \hspace{1.5in}
{\bf (b)}}
 \caption{{\bf (a)}\,\, Numerical evaluation of
$\zeta/\eta$ both on the big-BH branch (the solid curve) and on
the small BH branch (the dashed curve). $T_m$ denotes $T_{min}$.
{\bf (b)} \,\, The two branches of black-hole solutions, that
correspond to different ranges of $\l_h$. The big BH corresponds
to $\l_h<\l_{min}$ and the small BH corresponds to
$\l_h>\l_{min}$.}
 \label{illus}
 \end{figure}

 From the practical point of view, we find the second numerical method above (solving the fluctuation equation at $\omega=0$)
 easier in the range of $\l_h$ that corresponds
to the small black-hole. The result is shown in figure \ref{illus}
(a). The presence of two branches for $T>T_{min}$, is made clear
in this figure. See also fig \ref{illus} (b) for the respective ranges of
$\l_h$ that correspond to small and big BHs. In fig \ref{illus}
(a), $\zeta/s$ on the big BH branch is depicted with a solid curve
and the small BH branch is depicted with a dashed curve. We
observe that $\zeta/s$ keeps increasing on the big-BH branch as T
is lowered, up to the temperature $T_{min}$ where the small and
big BH branches merge\footnote{As far as the thermodynamics of the
gluon plasma is concerned, the temperatures below $T_c$ (on the
big BH branch) has little importance, because for $T<T_c$
the plasma is in  the confined phase.}. On the other hand, on the
small BH branch $\z/s$ keeps increasing as the T is increased, up
to a certain $T_{max}$ that lies between $T_{min} $ and $T_c$, see
figure \ref{bulkviscinset}. From this point onwards, $\z/s$
decreases with increasing T.

\begin{figure}
 \begin{center}
\includegraphics[height=6cm,width=9cm]{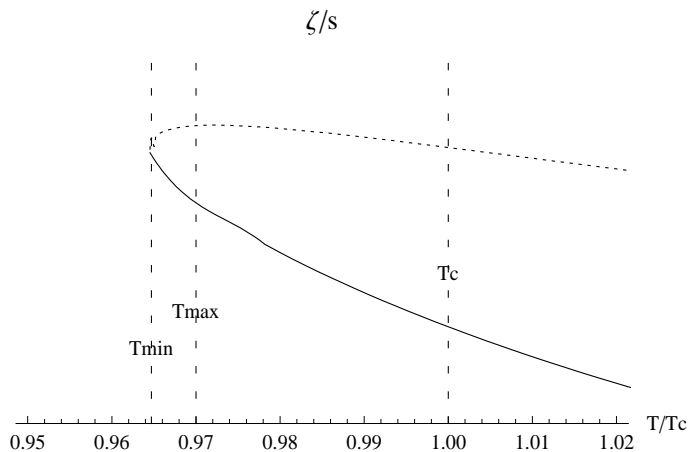}
 \end{center}
 \caption[]{An inset from the figure \ref{illus} around the maximum of $\zeta/s$. }
\label{bulkviscinset}
 \end{figure}

A simple fact that can be proved analytically is that the
derivative of $\zeta/s$ diverges at $T_{min}$. This is also clear
from figure \ref{bulkviscinset}. This is shown by inspecting
 equation (\ref{zs}). The T derivative is determined as
$d/dT = (dT/d\l_h )d/d\l_h$. Whereas the derivative w.r.t $\l_h$
is everywhere smooth\footnote{Note that $c_b$ is also a function
of $\l_h$. As both the fluctuation equation (\ref{hw}) and the
boundary conditions are smooth at $\l_h=\l_{min}$, one concludes
that $c_b$ also is smooth at this point.}, the factor $dT/d\l_h$
diverges at $T_{min}$ by definition, see figure \ref{illus} (b).

Therefore, we propose that the presence of a $T_{min}$ where the
big and the small black-holes meet, in other words, presence of a
small-black-hole branch is  responsible for the increase of $\z/s$
near $T_{min}$. As in most of the holographic constructions that
we analyzed, and specifically in the example we present in this
paper, $T_c$ and $T_{min}$ are close to one another, this fact
implies a rise in the bulk viscosity near $T_c$. This proposal,
combined with the fact that {\em the existence of a small BH
branch and color confinement in the dual gauge theory at zero T
are in one-to-one correspondence} \cite{gkmn2}, suggests that {\em
in confining large-N gauge theories, there will be a peak in the
ratio $\zeta/s$ close to $T_c$.}

Another fact that can be shown analytically is that $\zeta/s$
asymptotes to a finite value as $T\to\infty$ in the small
black-hole branch \footnote{See equations (\ref{fe1}),
(\ref{fe2}), and the discussion in
 App. B.}. We find that,
\be\lab{sbhlim} \frac{\zeta}{s}\bigg|_{small}
\to \frac{1}{6\pi}, \qquad as\,\,\, T\to\infty.
 \ee
As the entropy density vanishes in this limit \cite{gkmn2}, we
conclude that $\zeta$ should vanish with the same rate.

For a general potential with strong coupling asymptotics \be
V(\l)\sim \l^{Q}~~~~~{\rm as}~~~~\l\to\infty, \ee taking into
account (\ref{zs}), equation (\ref{sbhlim}) is modified to
\be\frac{\zeta}{s}\bigg|_{small} \to \frac{3Q^2}{32\pi}, \qquad
as\,\,\, r_h\to r_0. \label{an1} \ee where $r_0$ is the position
of the singularity in the zero temperature solution.

For confining
theories, the limit $r_h\to r_0$ corresponds to $T\to\infty$ on
the small BH branch. However, one can show that the result
(\ref{an1}) holds quite generally, regardless of whether the zero T theory
confines or not\footnote{The arguments in App. B remain valid in
the general case.}. In particular, for the non-confining
theories---that is either when $Q<4/3$ or when $Q=4/3$ but the
subleading term in the potential vanishes at the singularity---there
is only the big black-hole branch and the limit $r_h\to r_0$
corresponds to the zero T limit of this BH. Thus, we also learn
that there exist holographic models that correspond to non-confining gauge theories
whose zero T limit yield a
constant $\zeta/s$. This constant approaches zero as $Q\to 0$, i.e.
in the limiting AdS case.

We also see that the asymptotic value of $\z/s$ in the small BH
branch is close to the value of $\z/s$ near $T_c$. We shall give
an explanation of this fact in the next subsection. Using the
asymptotic formula (\ref{an1}), the fact that $Q>{4\over 3}$ for
confinement and $Q\leq {4\sqrt{2}\over 3}$ for the IR singularity
to be good and repulsive we may obtain a range of values where we
expect $\z/s$ to vary, namely \be {1\over 6\pi}~~\leq~~
\frac{\zeta}{s}\bigg|_{\rm small, asymptotic}~~\leq~~ {1\over
3\pi}. \label{an2}
 \ee

A final observation concerns the coefficient $c_b(\l_h)$ in
(\ref{zs}). This part is the only input from the solution of the
fluctuation equation, the rest of (\ref{zs}) is  fixed by the
dilaton potential entirely. We plot the numerical result for $c_b$ in fig \ref{cblh}
as a function of the coupling at the horizon $\l_h$.

First of all, Figure \ref{cblh} provides a
check that, the approximate bound of \cite{Gubser:2008sz}
$|c_b|\geq1$, is satisfied in the entire range. One also observes
$c_b$ approaches to 1 in the IR and UV asymptotics. These facts can
be understood analytically: In the UV (near the boundary) it is
because of the boundary condition $c_b=1$.
In the IR, it is more
subtle, and we explain this in appendix \ref{AppB}.

Finally, we observe that the
deviation of $c_b$ from the asymptotic value 1 is maximum around
the phase transition point $\l_c$. In fact, we numerically
observed that the top of the curve in figure \ref{cblh} coincides
with $\l_c$ to a very high accuracy. Whether this is just a
coincidence or not, it needs to be clarified.

\begin{figure}
 \begin{center}
\includegraphics[height=6cm,width=9cm]{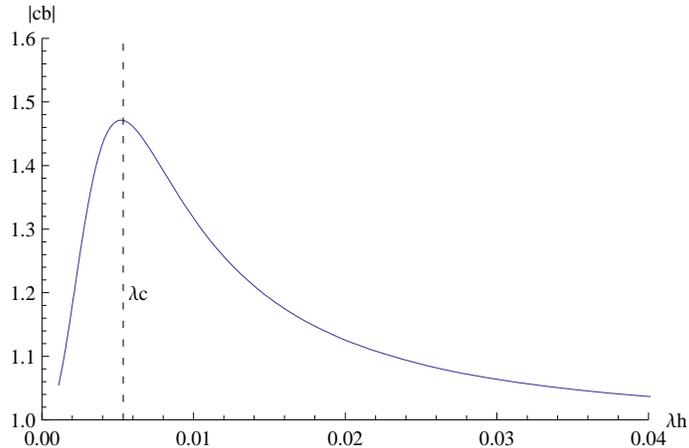}
 \end{center}
 \caption[]{The coefficient $|c_b|$ of equation (\ref{zs}) as a function of $\l_h$.}
\label{cblh}
 \end{figure}

\subsection{The adiabatic approximation}

Motivated by the Chamblin-Reall solutions \cite{CR}, Gubser et al.
\cite{GubserMimick} proposed an approximate adiabatic formula for
the speed of sound. In the case when $V'/V$ is a slowly varying
function of $\f$, \cite{GubserMimick} proposes the following
formulae for the entropy density and the temperature:
\bea \log s
&=& -\frac83\int^{\f_h} d\f \frac{V}{V'} + \cdots,\lab{adb1}\\
\log T &=& \int^{\f_h} d\f \le(\half \frac{V'}{V}
-\frac89\frac{V}{V'}\ri) \cdots,\lab{adb2}
\eea
where ellipsis
denote contributions slowly varying in $\f_h$\footnote{Various
coefficients in these equations  differ from \cite{GubserMimick}
due to our different normalization of the dilaton kinetic term.}.

\begin{figure}
 \begin{center}
\includegraphics[height=6cm,width=9cm]{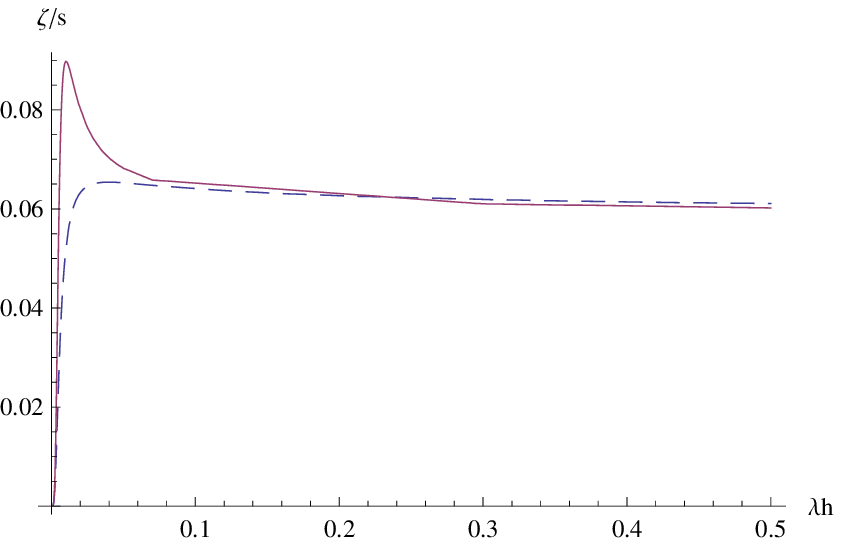}
 \end{center}
 \caption[]{Comparison of the exact $\zeta/s$ with the adiabatic approximation in the variable $\l_h$.
 Solid(red) curve is the full numerical result and the dashed(blue) curve follows from (\ref{zsadb}).}
\label{figadb1}
 \end{figure}

It is very useful to reformulate this approximation using the
method of scalar variables, which in turn allows us to extract the
general T dependence of most of the thermodynamic observables in
an approximate form. Here, we apply this formalism to the
computation of $\zeta/s$. We explain the method of scalar
variables in Appendix {\ref{AppA} and the details of the adiabatic
approximation in the scalar variables are given in Appendix
\ref{AppC}.

For the scalar variable $X$ (see Appendix \ref{AppA}
for a definition), the adiabatic approximation means
\be\lab{Xadb}
X(\f) \approx -\frac38 \frac{V'(\f)}{V(\f)}.
\ee
In Appendix
\ref{AppC} we present an independent argument based on the
Einstein's equations in scalar variables, for why this
approximation holds in certain regimes. The fluctuation equation
(\ref{hw}) greatly simplifies with (\ref{Xadb}). In fact, as shown
in Appendix \ref{AppC}, the solution becomes independent of $\f$. With unit
normalization on boundary, the adiabatic solution in the entire
range of $\f\in(-\infty, \f_h)$ becomes $h_{adb}(\f)=1$.
Consequently, the coefficient $c_b$ in (\ref{zs}) becomes unity,
hence:
\be
 \label{zsadb} \frac{\zeta}{s}\bigg|_{adb} =
\frac{3}{32\pi} \left(\frac{V'(\f_h)}{V(\f_h)}\right)^2.
\ee

We
plot this function in $\l_h$ in figure \ref{figadb1}, where we also
provide the exact numerical result for comparison. Note that in figure \ref{figadb1}
the whole large black-hole branch has been compressed at the left of the figure for $\l_h\lesssim 0.04$
The same
functions in the variable $T/T_c$ are plotted in figure \ref{figadb2}.

\begin{figure}
 \begin{center}
\includegraphics[height=6cm,width=9cm]{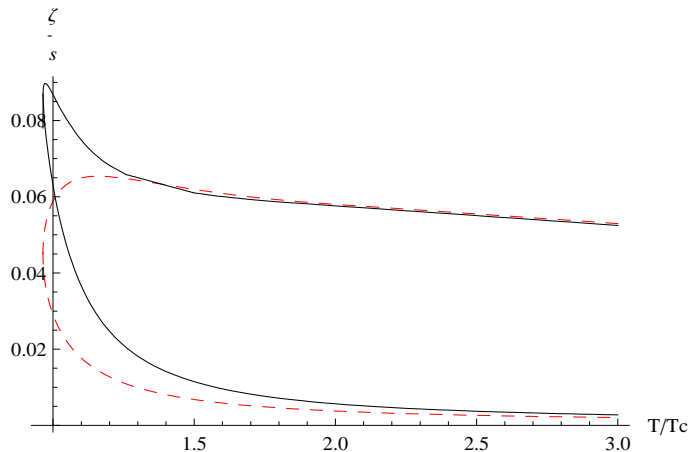}
 \end{center}
 \caption[]{Comparison of the exact $\zeta/s$ with the adiabatic approximation in variable T. Solid(blue) curve is the full numerical
 result and the dashed(red) curve follows from (\ref{zsadb}).}
\label{figadb2}
 \end{figure}

The validity of the adiabatic approximation equation (\ref{Xadb}), is
determined by the rate which $V'/V$ varies with $\phi$.  In
particular, the approximation becomes exact in the limits where
$V'/V$ becomes constant. This happens for a constant potential or a potential that is a single power of $\l$ (eponential in $\phi$).
This is the case in the UV
($\f\to-\infty$, where the potential becomes a constant) and the IR ($\f\to + \infty$ where the potential becomes a power law.)
. Therefore
equation (\ref{zsadb}) allows us to extract the analytic behavior of
$\zeta/s$ in the limits $\phi_h\to\pm\infty$.

 The numerical values
one obtains from (\ref{zsadb}) in the intermediate region may
differ from the exact result (\ref{zs}) considerably, especially
near $T_c$. However, we expect that the general shape will be
similar. We refer to Appendix \ref{AppC} for further  details.

Finally, the adiabatic approximation hints at why, in the
particular background that we study, $\zeta/s$ at $T_c$ is close
to the limit value (\ref{sbhlim}): In order to see this we rewrite (\ref{zsadb})
as \be \label{zsadb2} \frac{\zeta}{s}\bigg|_{adb} =
\frac{2}{3\pi} X^2. \ee In the limit (\ref{sbhlim}) we have
 $X\to -1/2$. The only other point where $X = 1/2$, is at the
 minimum of the string frame scale factor $\f_*$. This is the
 point where the confining string saturates \cite{ihqcd2}.
On the other hand, we expect on general physical grounds that the
de-confinement phase transition happens near this point, i.e.
$\f_c\approx \f_*$. Thus, the adiabatic formula predicts that
$\zeta/s (\f_c)$ be close to the limit value
$1/6\pi$.\footnote{This argument may break down for two (dependent) reasons:
First of all the adiabatic approximation becomes lees good
near $\f_c$. This is because, in this region $V'/V$ varies
relatively more rapidly as a function of $\f$. Secondly, precisely
because of this, even though $\f_c$ is not far away from $\f_*$
the difference can result in a considerable change in the value of
$\zeta/s$ through (\ref{zsadb}).}

\begin{figure}[h!]
 \begin{center}
\includegraphics[height=6cm,width=9cm]{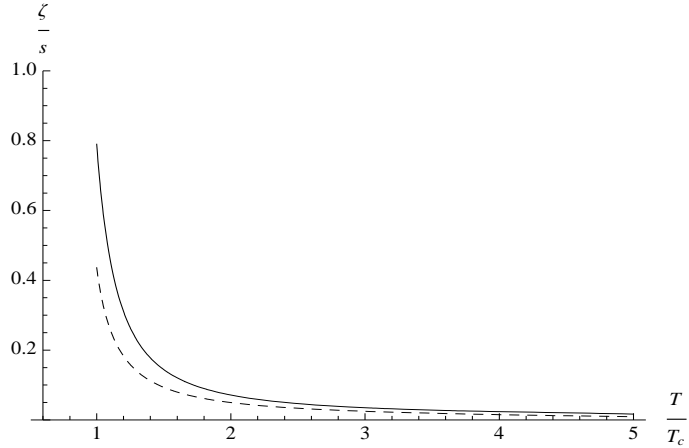}\\
(a)\\
\includegraphics[height=6cm,width=9cm]{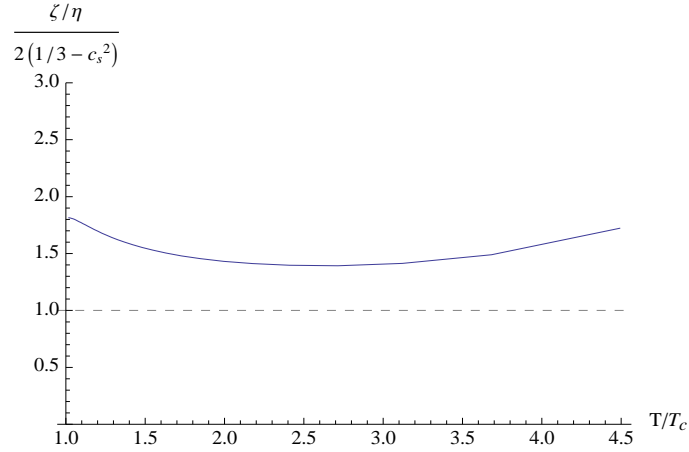}\\
(b)
 \end{center}
 \caption[]{(a) Comparison of $\zeta/\eta$ (solid line) and the RHS of (\ref{buchel1}) (dashed line),
obtained using the speed of sound of the IHQCD model \cite{gkmn3}.
(b) Plot of  the function $C(T)$ defined in equation (\ref{buchel2}) as a
function of temperature. The horizontal dashed line indicates where
Buchel's bound is saturated.
 We see that the bound is satisfied in the entire range of temperatures.}
\label{buchelfig}
 \end{figure}

\subsection{Buchel's bound}

In \cite{Buchel}, Buchel proposed a bound  for the ratio of the bulk and shear viscosities,
 motivated by certain well-understood holographic examples.
In 4 space-time dimensions the Buchel bound reads,
\be\label{buchel1}
\frac{\zeta}{\eta} \geq 2\left(\frac13 -c_s^2\right).
\ee
We note that the bound is proposed to hold in the entire range of temperature from $T_c$ to $\infty$.
This bound is trivially satisfied for exact conformal theories such as ${\cal N}=4$ YM, and
saturated in theories on $Dp$ branes \cite{Buchel,kas}.
With the numerical evaluation at hand, we can check (\ref{buchel1}) in our case. In figure \ref{buchelfig} (a)  we plot the LHS and RHS of the bound
\footnote{Since this theory contains two derivatives only, ${\eta\over s}$ has the
 universal value $1/4\pi$.}. We clearly
see that the bound is satisfied for all temperatures. As expected, both the LHS and the RHS of (\ref{buchel1}) vanishes in the high T conformal limit.

A clear picture of Buchel's bound is obtained by defining the function:
\be\label{buchel2}
C(T) = {\zeta/\eta \over 2\left(1/3 -c_s^2\right)},
\ee
in terms of which the bound is simply $C > 1$. In Figure \ref{buchelfig} (b)
 we show the function $C(T)$  obtained numerically in our  IHQCD model, between $T_c$  and  $5T_c$.
  The values of this function are mildly dependent
on temperature, and are between 1.5 and 2, the same range of values that were recently considered in the hydrodynamic codes
by Heinz and Song \cite{Heinz}.

% \begin{figure}[h!]
%  \begin{center}
% \includegraphics[height=6cm,width=9cm]{BuchelBound2905.eps}
%  \end{center}
%  \caption[]{Comparison of $\zeta/\eta$ (solid line) and the RHS of (\ref{buchel1}) (dahsed line),
% obtained using the speed of sound of the HIQCD model \cite{gkmn3}.
%  We see that the bound is satisfied in the entire range of temperatures.}
% \label{buchelfig}
%  \end{figure}

\section{The drag force on strings and heavy quarks}\label{drag}

We will now consider an (external) heavy quark moving through an infinite volume of  gluon plasma with a fixed velocity $v$ at a finite temperature T
\cite{her,gub2}.
  The quark feels a drag force coming from its interaction with the plasma
  and an external force has to be applied in order for it to keep a constant velocity.
  In a more realistic set up one would like to describe the deceleration caused by the drag.

  The heavy external quark can be described by a string whose endpoint is at the boundary.
  One can accommodate flavor by introducing D-branes, but we will not do this here.
  A first step is to describe the classical string ``trailing" the quark.

  We consider the Nambu-Goto action on the worldsheet of the string.
\begin{equation}\label{NGACTION}
S_{NG} = -\frac{1}{2\pi \ls^2}\int d\sigma d\tau \sqrt{det\left(-g_{MN} \partial_{\alpha}X^{M} \partial_{\beta}X^{N}\right)} \;,
\end{equation}
where the metric is the string frame metric.
The ansatz we are going to use to describe the trailing string is, \cite{gub1},
\begin{equation}
\label{TRAILANSATZ}
 X^{1} = v t +\xi(r),\quad X^{2}=X^{3}=0 \;,
\end{equation}
along with the gauge choice
\begin{equation}\label{TRAILGAUGE}
\sigma =r, \quad \tau =t\;,
\end{equation}
where $r$ is the (radial) holographic coordinate. The string is moving in the $X^1$ direction.

This is a ``steady-state" description of the moving quark as acceleration and deceleration are not taken into account.
  For a generic background the action of the string becomes
\begin{equation}
S = -\frac{1}{2\pi \ls^2} \int dt dr ~\sqrt{-g_{00}g_{rr}-g_{00}g_{11}\xi'^{2}-g_{11}g_{rr}v^2} \;.
\end{equation}
Note that $g_{00}$ is negative, and we should check whether our solution produces a real action.
For example a straight string stretching from the quark to the horizon is a solution to the equations of
 motion but has imaginary action.

 We note that the action does not depend on $\xi$ but only its derivative, therefore the corresponding ``momentum" is conserved
\begin{equation}\label{CONSMOMENTUM}
\pi_{\xi} = -\frac{1}{2\pi \ls^2} \frac{g_{00}g_{11}\xi'}{\sqrt{-g_{00}g_{rr}-g_{00}g_{11}\xi'^2-g_{11}g_{rr}v^2}}\;.
\end{equation}
We solve  for $\xi'$ to obtain
\begin{equation}\label{GOTMOMENTUM}
\xi' = \frac{\sqrt{-g_{00}g_{rr}-g_{11}g_{rr}v^2}}{\sqrt{g_{00}g_{11}\left(1+g_{00}g_{11}/(2\pi\ls^2 \pi_{\xi})^2\right)}} \;.
\end{equation}
The numerator changes sign at some finite value of the fifth coordinate $r_{s}$.
 For the solution to be real, the denominator has to change sign at the same point.  We therefore determine $r_{s}$ via the equation
\begin{equation}\label{GOTZS}
 g_{00}(r_{s})+g_{11}(r_{s})v^2=0\;,
\end{equation}
and the constant momentum
\begin{equation}\label{GOTPIXI}
 \pi_{\xi}^{2} = -\frac{g_{00}(r_{s})g_{11}(r_{s})}{(2\pi \ls^2)^2}\;.
\end{equation}
Writing the string-frame metric as
\be
ds^2=e^{2A_s}\left[{dr^2\over f}-f~dt^2+dx\cdot dx\right]
\label{b1}\ee
(\ref{GOTZS}) becomes
\be
v^2=f(r_s)
\label{b3}\ee
The induced world-sheet metric is therefore
\be
g_{\a\b} = e^{2A_s(r)}\left( \begin{array}{cc}
-(f(r)-v^2)~~ & ~~{e^{2A_s(r_s)}v^2\over f(r)}\sqrt{f(r)-v^2\over  e^{4A_s(r)}f(r)-e^{4A_s(r_s)}v^2} \\
{e^{2A_s(r_s)}v^2\over f(r)}\sqrt{f(r)-v^2\over  e^{4A_s(r)}f(r)-e^{4A_s(r_s)}v^2}~~ & {e^{4A_s(r)}f^2(r)-v^4e^{4A_s(r_s)}\over
f^2(r)\left[e^{4A_s(r)}f(r)-v^2{e^{4A_s(r_s)}}\right]}
\end{array}\right)\label{indu}
\end{equation}

We can change the time coordinate to obtain a diagonal induced metric $t=\tau+\zeta(r)$ with
$$
\zeta'={e^{2A_s(r_s)}v^2\over f(r)\sqrt{(f(r)-v^2)( e^{4A_s(r)}f(r)-e^{4A_s(r_s)}v^2)}}
$$
The new metric is
\be
ds^2=e^{2A_s(r)}\left[-(f(r)-v^2)d\tau^2+{e^{4A_s(r)}\over (e^{4A_s(r)}f(r)-e^{4A_s(r_s)}v^2)}dr^2\right]
\end{equation}
and near $r=r_s$ it has the expansion
\be
ds^2=\left[-f'(r_s)e^{2A_s(r_s)}(r-r_s)+{\cal O}((r-r_s)^2)\right]d\tau^2+\left[{e^{2A_s(r_s)}\over
(4v^2 A'_s(r_s)+f'(r_s))(r-r_s)}+{\cal O}(1)\right]dr^2
\ee
This is a world-sheet black-hole metric with horizon at the turning point $r=r_s$.

\subsection{The drag force}

The drag force on the quark can be determined by calculating  the momentum that is lost by flowing along  the string into the horizon:
\begin{equation}\label{GOTDRAGFORCE}
F_{\rm drag} = \frac{dp_{1}}{dt}=-\frac{1}{2\pi\ls^2}\frac{g_{00}g_{11}\xi'}{\sqrt{-g}}=\pi_{\xi}\;.
\end{equation}
 This can be obtained by considering the world-sheet Noether currents $\Pi^{\alpha}_{M}$\cite{Lawrence:1993sg}
and expressing the loss of momentum  as $\Delta P^{z}_{x_1} = \int\Pi^{r}_{1} $.
This may be  evaluated  at any
   value of r, but it is more convenient to evaluate it at $r=r_{s}$.

   We finally find that
\begin{equation}\label{DRAGEXPL}
 F_{\rm drag} = - \frac{1}{2\pi \ls^2} \sqrt{-g_{00}(r_{s})g_{11}(r_{s})}\;.
\end{equation}

Using the form (\ref{b1}) of our finite-temperature metric in the string frame
we finally obtain
\be
 F_{\rm drag} = - \frac{e^{2A_s(r_s)}\sqrt{f(r_s)}}{2\pi \ls^2}=-\frac{ e^{2A(r_{s})}\sqrt{f(r_s)} \lambda(r_{s})^{4/3}}{2\pi\ls^2}\;,
\label{b2}\ee
where in the second equality we expressed the force
in terms of the Einstein-frame scale factor and the ``running" dilaton.
Substituting from (\ref{b3}) we obtain
\be
 F_{\rm drag} = - \frac{v~e^{2A_s(r_s)}}{2\pi \ls^2}=-\frac{ v ~e^{2A(r_{s})}\lambda(r_{s})^{4/3}}{2\pi\ls^2}\;,
\label{b4}\ee
Before proceeding further, we will evaluate the drag force for the conformal case of ${\cal N}=4$ SYM
where
\be
e^{A_s}={\ell\over r}\sp v^2=f(r_s)=1-(\pi T r_s)^4\sp {\ell^2\over \ls^2}=\sqrt{\l}
\label{b5}\ee
Substituting in (\ref{b4}) we obtain, \cite{her}-\cite{gub2},
\be
F_{\rm conf}={\pi\over 2}\sqrt{\lambda}~T^2{v\over \sqrt{1-v^2}}
\label{a9}\ee

Moving on to YM, to compute the drag force from equation (\ref{b4}) we must
first determine $\ls$ in the IHQCD model. In this setup there is no analog
of the ${\cal N}=4$ SYM relation (\ref{b5}) between $\ell$,  $\ell_s$ and $\l$.
Rather, the fundamental string length $\ls$ is determined in a bottom-up fashion, by matching the effective string tension to the QCD string tension $\sigma_c$
derived from the lattice calculations.  From (\ref{strten})
\begin{equation}
\sigma_c= \frac{1}{2\pi\ls^2} e^{2A_{s,o}(r_{*})} =\frac{1}{2\pi\ls^2} e^{2A_o(r_{*})} \lambda_o(r_{*})^{4/3}\;,
\label{b6}\end{equation}
where $r_{*}$ is the point where the zero-temperature string scale  factor $A_{s,o} (r)$ has a minimum.
 For a typical value of $\sigma_c\sim (440 \;MeV)^2$ \cite{chenetal} we find
\begin{equation}
 \ell_s = 0.15 ~\ell \;,
\label{b7}\end{equation} where $\ell$ is the radius of the
asymptotic AdS space.

 On the other hand, unlike in ${\cal N} = 4$ SYM,
in the IHQCD model the value of the coupling $\l(r_s)$ in equation (\ref{b4})  is not an extra parameter to be fixed by hand, but rather it is determined dynamically together with the background metric.

\begin{figure}[ht]
\centering
\includegraphics[width=10cm]{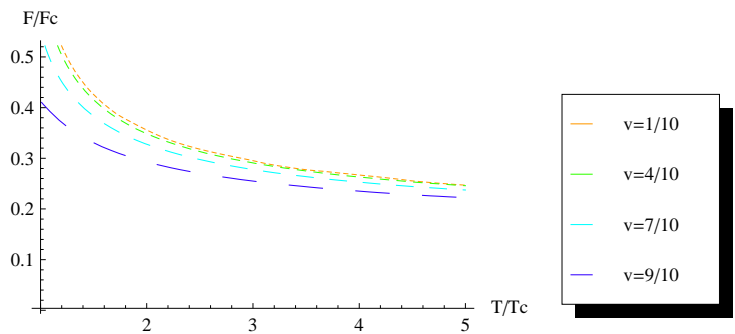}
\caption{In this figure the ratio of the drag force in improved
holographic QCD to the drag force in ${\cal N}=4$ SYM is shown.
The ratio is computed for different velocities as a function of
temperature. The 't Hooft coupling for the ${\cal N}=4$ SYM
theory is taken to be $5.5$. We chose this value as it is considered
in the central region of possible values for the 't Hooft coupling. It is seen that as the velocity increases the value of the ratio decreases.}\label{MyFdragFigureA}
\end{figure}

\begin{figure}[ht]
\centering
\includegraphics[width=10cm]{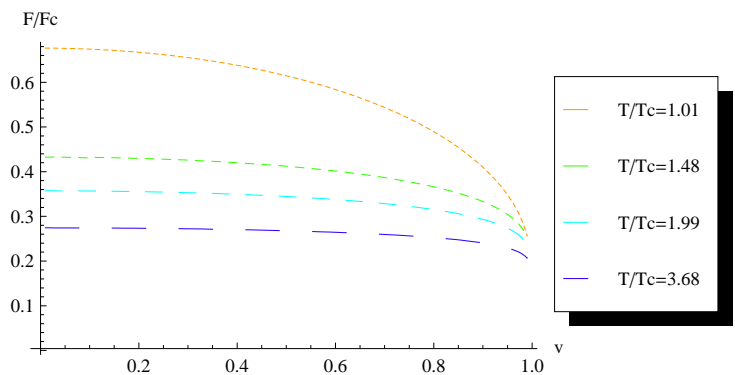}
\caption{In this figure the ratio of the drag force in improved holographic QCD to the drag force in ${\cal N}=4$ SYM is shown.
The ratio is computed for different temperatures as a function of velocity.
The 't Hooft coupling for the ${\cal N}=4$ SYM theory is taken to be $5.5$.  As temperature increases the value of the ratio decreases.}\label{MyFdragFigureB}
\end{figure}

\subsection{The relativistic asymptotics}

When $v\to 1$ then $r_s\to 0$ and we approach the boundary.
Near the boundary ($r\to 0$) we have the following asymptotics of the scale factor and the 't Hooft coupling, \cite{gkmn2}
\be
f(r)\simeq 1-{\pi T~e^{3A(r_h)}\over \ell^3}r^4\left[1+{\cal O}\left({1\over \log(\Lambda r)}\right)\right]+{\cal O}(r^8)
\sp e^{A(r)}={\ell\over r}\left[1+{\cal O}\left({1\over \log(\Lambda r)}\right)\right]+\cdots
\label{a2}\ee
and
\be
\lambda(r)=-{1\over \beta_0\log(r\Lambda)}+{\cal O}(\log(r\Lambda)^{-2})
\label{b8}\ee
where $r_h$ is the position of the horizon.

We therefore obtain for the turning point
\be
r_s\simeq \left[{\ell^3(1-v^2)\over \pi T e^{3A(r_h)}}\right]^{1\over 4}\left[1+{\cal O}\left({1\over \log(1-v^2)}\right)\right]
\sp \lambda(r_s)\simeq
-{4\over \beta_0 \log \left[1-v^2\right]} +\cdots
\label{a3}\ee
and the drag force
\be
F_{\rm drag}\simeq -{\sqrt{\pi T\ell b^3(r_h)\lambda^{8\over 3}(r_s)}\over 2\pi\ell_s^2}{v\over \sqrt{1-v^2}}+\cdots
\label{a4}\ee
We also use
\be
e^{3A(r_h)}={s(T)\over 4\pi M_p^3~ N_c^2}={45\pi\ell^3 s(T)\over N_c^2}
\label{a7}\ee
where $s(T)$ the entropy per unit three-volume, and we write the relativistic asymptotics of the drag force as,
\be
F_{\rm drag}\simeq -{\sqrt{\pi T\ell b^3(r_h)}\over 2\pi\ell_s^2}{v\over \sqrt{1-v^2}\left(-{\beta_0\over 4}\log\left[1-v^2\right]\right)^{4\over 3}}+\cdots
\label{b9}\ee
$$
=-{\ell^2\over \ell_s^2} \sqrt{{45 ~T s(T)}\over 4N_c^2}{v\over \sqrt{1-v^2}\left(-{\beta_0\over 4}\log\left[1-v^2\right]\right)^{4\over 3}}+\cdots
$$

The force is proportional to the relativistic  momentum
combination $v/\sqrt{1-v^2}$ modulo a power of
$\log\left[1-v^2\right]$. This factor is present because, as
argued in \cite{k} the asymptotic metric is AdS in the Einstein
frame instead of the string frame. Its effects are not important
phenomenologically. We discuss this issue further in Appendix \ref{UV}.

\subsection{The non-relativistic asymptotics}

We now consider the opposite limit,  $v\to 0$. In this case the turning point asymptotes to the horizon, $r_s\to r_h$
and we have the expansion
\be
f(r)\simeq 4\pi T(r_h-r)+{\cal O}((r_h-r)^2)\sp r_s=r_h-{v^2\over 4\pi T}+{\cal O}(v^4)
\label{a10}\ee
and
\be
F_{\rm drag}\simeq -{e^{2A(r_h)}\lambda(r_h)^{4\over 3}\over 2\pi\ell_s^2}v\left[1-{v^2\over 2\pi T}A'(r_h)-
{v^2\over 3\pi T}{ \lambda'(r_h)\over \lambda(r_h)}+{\cal O}(v^4)\right]
\label{a11}\ee
$$
\simeq -{\ell^2\over \ls^2}\left({45\pi ~s(T)\over N_c^2}\right)^{2\over 3}{\l(r_h)^{4\over 3}\over 2\pi}v+{\cal O}(v^3)
$$
where primes are derivatives with respect to the conformal coordinate $r$.
%The 'tHooft coupling constant at the horizon $\l_h\equiv \l(r_h)$ is plotted as a function of temperature in figure (\ref{HorizonDilaton})
%\begin{figure}[ht]
%\centering
%\includegraphics[width=10cm]{dilatonhorizonInter.eps}
%\caption{The value of the 't Hooft coupling constant at the horizon is shown as a function of the temperature.}\label{HorizonDilaton}
%\end{figure}

\subsection{The diffusion time}

For a  heavy quark with mass $M_q$ we may rewrite
(\ref{a9}) as
\be
F_{\rm conf}\equiv {dp\over dt}=-{1\over \tau}p\sp p={M_q v\over \sqrt{1-v^2}}
\label{b10}\ee
where the first equation defines the diffusion time $\tau$. In the conformal case, the diffusion time
is constant,
\be\label{b10-2}
\tau_{conf}={2M_q\over \pi \sqrt{\l}~T^2}
\ee
%The characteristic time $\tau$ is constant in this case.
This is not anymore the case in QCD, where $\tau$ defined as above is momentum dependent.
We may still define it as in (\ref{b10}) in which case we obtain the following limits
\be
\lim_{p\to\infty}~\tau=M_q~{\ls^2\over \ell^2} \sqrt{4N_c^2\over 45 ~T s(T)}
\left({\beta_0\over 4}\log{p^2\over M_q^2}\right)^{4\over 3}+\cdots
\label{b11}\ee
\be
\lim_{p\to 0}~\tau=M_q~{\ls^2\over \ell^2}\left({N_c^2\over 45\pi ~s(T) }\right)^{2\over 3}{2\pi\over \l(r_h)^{4\over 3}}+\cdots
\label{b12}\ee

\begin{figure}[h!]
\centering
\includegraphics[width=13cm]{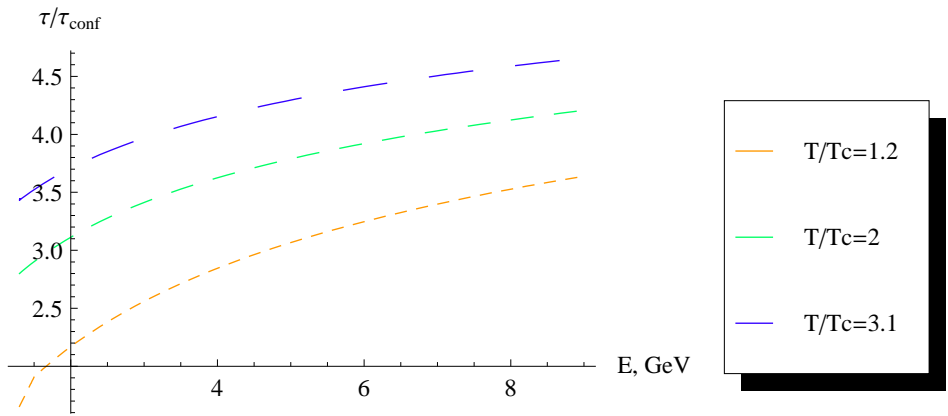}
\caption[]{
In this figure the ratio of the diffusion time in the Improved Holographic QCD model to the diffusion time in ${\cal N}=4$ SYM is shown.
 The 't Hooft coupling for ${\cal N}=4$ SYM is taken to be $\lambda=5.5$.
 The heavy quark has a mass of $M_{q}=1.3 GeV$.
  Note that with the definition of the diffusion time in (\ref{b10}) the ratio is the inverse of the ratio of the forces.
   A similar plot is valid for the bottom quark as well, as the mass drops out of the ratio.
     although the energy scales are different.
      In this plot the x-axis is taken to be in MeV units. As temperature increases the ratio also increases.
}\label{TDiffusionA}
\end{figure}

\subsection{Including the medium-induced  correction to the quark mass\label{mass}}

In order to estimate the diffusion time of a quark of finite  rest
mass, we must take into account the fact that the mass of the
quark receives medium-induced  corrections. In other words, the mass
appearing in equation (\ref{b10}) is a temperature-dependent quantity,
$M_q(T) \neq M_q(T=0)$. The ratio $M_q(T)/M_q(0)$ can be estimated
holographically by representing a static quark of finite mass by a
static, straight string \footnote{This representation ignores the
fact that the {\em kinetic} mass of a moving quark may be
different from the static mass \cite{her}. We plan to treat this
in the future.} stretched along the radial direction starting at a
point $r=r_q \neq 0$. At zero temperature, the IR endpoint of the
string can be taken as the ``confinement'' radius, $r_*$, where
the string frame metric reaches its minimum value; At finite
temperature, the string ends in the IR  at the BH
horizon\footnote{ It would stop at the confinement radius if the
latter were closer to the boundary than the horizon, i.e. if
$r_*(T) < r_h(T)$. However, in the model we are considering, in
the big BH branch we find that the relation $r_h < r_* $  is
always satisfied.}. The masses of the quark at zero and finite $T$
are related to the worldsheet action evaluated on the static
solution $(\tau = t, \sigma = r)$ :

\be \label{qmass}
M_q(0)  = {\ell \over 2\pi \ell_s^2} \int_{r_q}^{r_*} d r\, e^{2A_o(r)} \l_o^{4/3}(r)\, , \qquad   M_q(T)  = {\ell\over 2\pi \ell_s^2}  \int_{r_q}^{r_h} dr \,  e^{2A(r)} \l^{4/3}(r)\, .
\ee

The value $r_q$ can be fixed numerically  by  matching $M_q(0)$ to the physical quark mass, and translating
the fundamental string tension  in physical units by using the relation (\ref{b6}), with $\sigma_{c} = (440 MeV)^2$. This makes $M_q(T)$ a function of $M_q(0)$. The ratios $M_q(T)/M_q(0)$ we found numerically in the model
under consideration  is shown in figure \ref{quarkmass} for the Charm ($M(0)=1.5 GeV$)  and Bottom ($M(0)=4.5 GeV$) quarks. The fact that, in
the deconfined plasma, the quark mass decreases with increasing temperature is a direct consequence of the holographic framework\footnote{For a possible
field theoretical explanation  of  this phenomenon, see \cite{marquet}.}, since for higher temperature, the distance to the horizon is smaller.
An indication  that this result may be  in the right direction
comes from the lattice computation of   the
shift in  the position of the quarkonium resonance peak at finite temperature \cite{Datta}:  in the deconfined phase the charmonium
 peak moves to lower mass at higher temperature. Our result for the
medium-induced shift in the constituent quark mass is consistent with these
observations.

\begin{figure}[ht]
\centering
\includegraphics[scale=1.2]{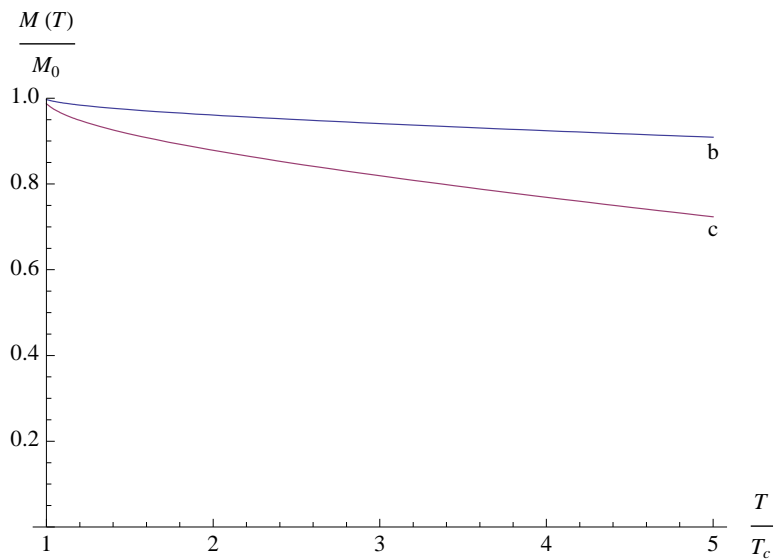}
\caption{Ratios between the thermal mass and the rest mass of the Charm (curve labeled ``c'') and Bottom (curve labeled ``b'' ) quarks, as a function of temperature.}\label{quarkmass}
\end{figure}

We can now  write the diffusion time from eqs. (\ref{b4}) and (\ref{b10}) as:
\be\label{stopt}
\tau(T,v) = {M_q(T) \over \sigma_{c} \sqrt{1-v^2}} \left({\l_o(r_*) \over \l(r_s)} \right)^{4/3}  e^{2A_o(r_*) - 2A(r_s)},
\ee
where once again we have eliminated the fundamental string length using equation (\ref{b6}). Given a set of zero- and
finite-temperature solutions, equation (\ref{stopt}) can be evaluated numerically for different values of the velocity and different quark masses. The results for the Charm ($M_q(0)=1.5 \; GeV$) and Bottom ($M=4.5 \;GeV$) quarks are displayed in figure \ref{stoppingF}.
\begin{figure}[ht]
\centering
\includegraphics[scale=0.8]{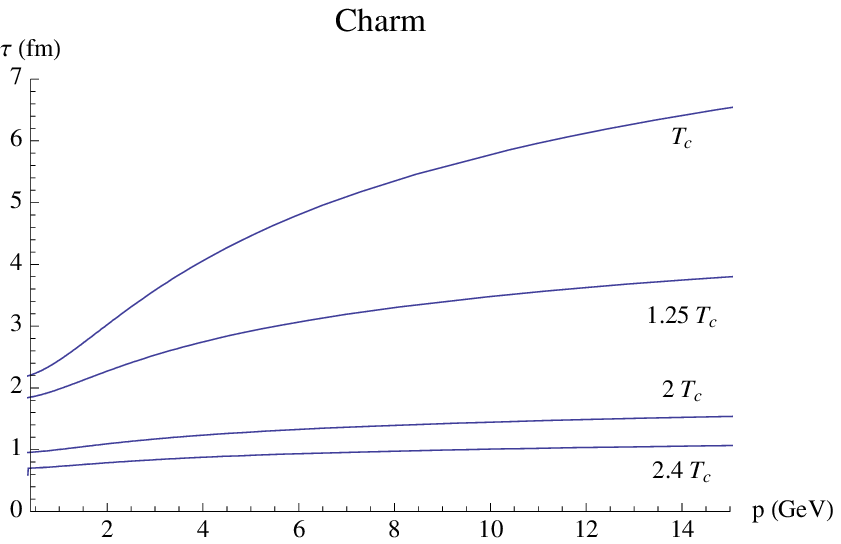}
\includegraphics[scale=0.8]{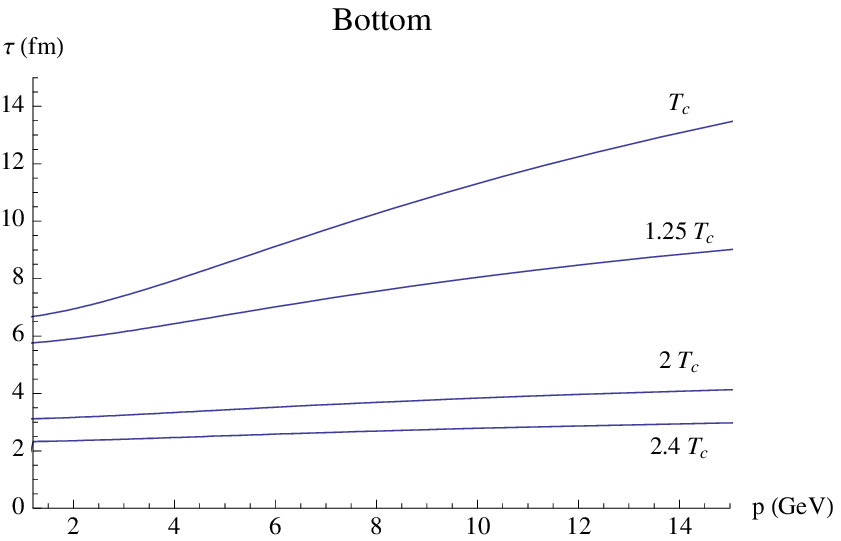}
\caption{Diffusion time for the Charm and Bottom quarks, as a function of energy, for different ratios of the temperature to the IHQCD transition temperature $T_c$.  }\label{stoppingF}
\end{figure}

\subsection{Temperature matching and diffusion time estimates}
An important question is how we should choose the  temperature in our holographic model in
order to compare our results with heavy-ion collision experiments. This is nontrivial, since
our setup is designed to describe pure $SU(N_c)$ Yang-Mills, whereas at RHIC temperatures
there are 3 light quark flavors that become relevant. As a consequence, the critical temperatures
and the number of degrees of freedom of the two theories are not  the same:
for pure $SU(N_c)$ Yang Mills we have $N_c^2 - 1$ degrees of freedom and a critical temperature around $260$ $MeV$; For $SU(N_c)$  QCD
with $N_f$ flavors the number of degrees of freedom is $ N_c^2 - 1  + N_c N_f$, and the transition temperature is lower, around  $180$ MeV. In our holographic
model, the transition temperature in physical units was estimated to be
$T_c =247$ MeV \cite{gkmn3}, i.e. close to the lattice result for the pure YM  deconfining temperature. From now on, this is the
value  we will mean  when  we refer to $T_c$. This is also close to the
temperature of   QGP at RHIC, which we will denote $T_{QGP}$, and  is estimated to be  around $250$ $MeV$.
Since this value is uncertain,  below we  give our results for a range of
temperatures between 200 MeV and 400 MeV. The higher temperatures
will be relevant for the LHC ion collision experiments (see e.g. \cite{lr3}).

Based on these considerations,  there are  different ways of fixing the temperature  (see  e.g.
the recent review \cite{gubser-review}): one {\em direct}  and
two {\em alternative} schemes (that we call the  {\em energy}  and {\em
entropy} scheme).
\begin{itemize}
\item {\bf{\em Direct scheme}}: The temperature of the holographic
model  is
identified with the temperature of the QGP in the experimental situation
(at RHIC or LHC),  $T_{ihqcd}^{(dir)} = T_{QGP}$.
 %  are identified. In our model, the overall energy scale
% was determined by the requirement that the critical temperature match the pure YM value of $T_c \simeq 260$ $MeV$.
% Thus, the comparison of  our model with  QGP  data  should be made  at a temperature $T_{ihqcd} = T_{QGP}$,
%in other words  $T_{ihqcd} / T_c \simeq 1$.
%{\bf\Large what is the critical temperature here? there was a
%remark before that is commented out that $T_c$ is matched to $260
%MeV$ in our model. But in the numerics paper we fixed the
%parameters by matching the latent heat not $T_c$, so we had $247
%MeV$ instead. I think we should state it clearly}

\item {\bf{\em Energy scheme}}: One matches the energy densities, rather than the temperatures. The energy
density at RHIC is approximately (treating the QCD plasma as a free gas\footnote{This is itself
an approximation, since as we know both from experiment and in our holographic model,  the plasma is strongly
coupled up to  temperatures of a few $T_c$}.)  $\epsilon_{QGP} \simeq (\pi^2/15) (N_c^2 - 1 + N_c N_f) (T_{QGP})^4$. For $N_c=N_f=3$, asking that our
energy density matches this value requires us to consider the holographic model at temperature $T_{ihqcd}^{(\epsilon)}$ given by
\be\label{altscheme}
\epsilon_{ihqcd} (T_{ihqcd}^{(\epsilon)}) \simeq 11.2 (T_{QGP})^4 %\qquad \Rightarrow \qquad T_* \simeq 1.25 \, T_c,
\ee
%as can be obtained numerically from the analysis of the thermodynamics performed in \cite{gkmn3}.
\item {\bf{\em Entropy scheme}}: Instead of matching the
energy densities, alternatively one can match the entropy density $s$,
which for the QGP, in the free gas approximation, is given   by
 $\s_{QGP} \simeq 4 \pi^2/45 (N_c^2 - 1 + N_c N_f) (T_{QGP})^4$. This leads
to the identification:
\be\label{entscheme}
s_{ihqcd} (T_{ihqcd}^{(s)}) = 14.9 (T_{QGP})^3
\ee
\end{itemize}

\begin{table}[h!]
\centering
\begin{tabular}{||c|c||c|c||c|c||}
\hline
$T_{QGP}$ (MeV) & $T_{QGP}/T_c$ & $T_{ihqcd}^{(\epsilon)}$ (MeV) &
$ T_{ihqcd}^{(\epsilon)}/T_c$  & $T_{ihqcd}^{(s)}$ (MeV) & $ T_{ihqcd}^{(s)}/T_c$ \\
\hline
190 & 0.77 &259 & 1.05  & 274& 1.11  \\
\hline
220 & 0.89 &290  & 1.18 & 302 & 1.23  \\
\hline
250 & 1.01 &325 & 1.31 & 335  & 1.35\\
\hline
280 & 1.13 &361 &1.46 & 368  & 1.49 \\
\hline
310 & 1.26 &398  &1.61 & 402 & 1.63 \\
\hline
340 & 1.38 &434  & 1.76 & 437 & 1.77 \\
\hline
370 & 1.50 &471  & 1.90 & 472 & 1.91 \\
\hline
400 &1.62  & 508  & 2.06 & 507 & 2.05  \\
\hline
%500 & 632 & & 2.43 &  \\
%\hline
\end{tabular}
\caption{Translation table between different temperature identification schemes. The first two columns display temperatures in the direct scheme, (in which the temperature of the holographic model matches the physical QGP temperature)
and the corresponding ratio to the IHQCD critical temperature,  that was
fixed by YM lattice results at $T_c = 247$ MeV \cite{gkmn3}.
The third and fourth columns  display the corresponding temperatures (and respective ratios to $T_c$)  in the energy scheme, and the last two in the  entropy scheme.
}\label{translation}
\end{table}

\begin{figure}[h!]
\centering
\includegraphics[scale=1.2]{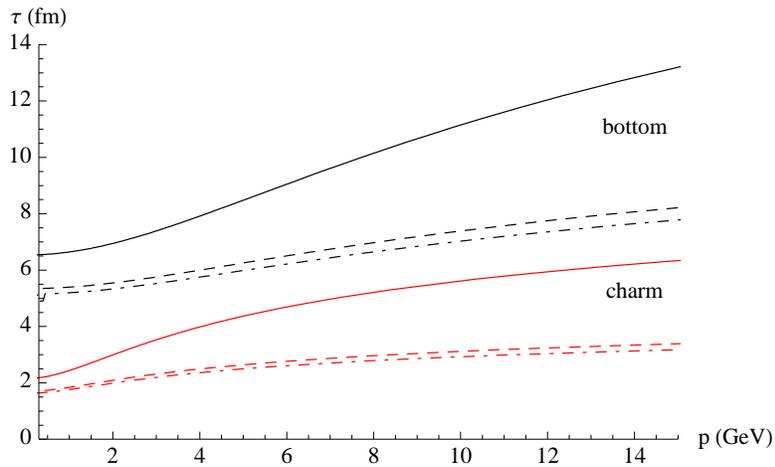}
\caption{Diffusion times for the Charm and Bottom quarks, as a function of
initial momentum, at $T_{QGP}=250 $ MeV. The different lines  represent the
 in the {\em direct} scheme (solid), {\em energy} scheme (dashed) and {\em entropy} scheme (dash-dotted), all corresponding to the same temperature $T_{QGP}=250 $ MeV.}\label{stopping2}
\end{figure}

\begin{table}[h!]
\centering
\begin{tabular}{|c|c c c|}
\hline
$T_{QGP},MeV$ & $\tau_{charm}$ (fm/c)  & $\tau_{charm}$ (fm/c)  &  $\tau_{charm}$   (fm/c )  \\
& (direct) & (energy) & (entropy) \\
\hline
220  & -  & 4.0  & 3.6   \\
\hline
250 & 5.7 &  3.1 & 3.0 \\
\hline
280 & 4.3   & 2.6  & 2.5  \\
\hline
310 & 3.5 &  2.1 &  2.1  \\
\hline
340 & 2.9 & 1.8 & 1.8 \\
\hline
370 & 2.5 & 1.5   &1.5 \\
\hline
400 & 2.1 &  1.3  & 1.3  \\
% \hline
% 500 & & 0.80 fm & 2.56 fm \\
\hline
\end{tabular}
\caption{The diffusion times for the charm quark are shown for
different temperatures, in the three different schemes.  Diffusion
times have been evaluated with a quark initial momentum  fixed  at
$p \approx 10$ $GeV$. }\label{DiffTablecharm}
\end{table}

\begin{table}[h!]
\centering
\begin{tabular}{|c|c c c|}
\hline
$T_{QGP} (MeV)$ & $\tau_{bottom}$ (fm/c)  & $\tau_{bottom}$ (fm/c)&  $\tau_{bottom}$ (fm/c)  \\
& (direct) & (energy)& (entropy)\\
\hline
220  & -  & 8.9 & 8.4 \\
\hline
250 & 11.4  & 7.5  & 7.1 \\
\hline
280 &  10.1 &  6.3 & 6.1\\
\hline
310 & 8.6  & 5.4  & 5.3 \\
\hline
340 & 7.5 & 4.7 & 4.7  \\
\hline
370 & 6.6 & 4.1  & 4.1  \\
\hline
400 & 5.8 & 3.6 & 3.6\\
% \hline
% 500 & & 0.80 fm & 2.56 fm \\
\hline
\end{tabular}
\caption{Diffusion times for the  bottom quark are shown for different temperatures, in the three different schemes.  Diffusion times have been evaluated with a quark initial momentum  fixed  at $p \approx 10$ $GeV$. }\label{DiffTablebottom}
\end{table}

The temperature translation table between  the various schemes is shown
in Table \ref{translation}. In that  table, $T_c = 247 MeV $ is the  deconfining  temperature of the holographic model.

In Figure \ref{stopping2} we show the
comparison between the diffusion times, as a function of initial  quark momentum,  in the different  schemes for
the Charm and Bottom quarks, at the  temperature $T_{QGP} = 250 MeV$.

 The results  for the diffusion times at different temperatures, computed at  a reference heavy quark initial momentum
$p \approx 10$  $GeV$,  are displayed in Tables \ref{DiffTablecharm} and \ref{DiffTablebottom}. We see that there is little practical difference between the {\em entropy}  and {\em energy} schemes; on the other hand  the difference between the {\em direct} scheme and the two alternative schemes
can be quite substantial.

%\subsection{Diffusion Constant}
%A diffusion constant can be defined via a generalized Einstein relation.  The diffusion constant will not be a constant here but will depend on momentum or energy.  This relation is derived in appendix \ref{langevin} and we reproduce it here
%\def\e{\varepsilon}\be
%D(E)={2MT\over \tau(M)}e^{E-M\over T}-2e^{E\over T}\int_{M}^{E}e^{-{\e\over T}}{\e d\e\over \tau(\e)}\sp E=\sqrt{p^2+M_q^2}
%\label{dd10}\ee
%For a constant diffusion time it becomes
%\be
%D(E)={2T\over \tau(M)}\left[E+T-Te^{E-M\over T}\right]
%\label{b13}\ee
%In the non-relativistic limit it gives the standard Einstein relation
%\be
%D\simeq {2MT\over \tau(M)}+{\cal O}(E_k^2)
%\label{b14}\ee
%with $E_k={p^2\over 2M}$ the non-relativistic kinetic energy.
%
%The relativistic diffusion coefficient for ${\cal N}=4$ sYM can be calculated from (\ref{b13}) with
%$\tau(M)$ given in (\ref{b10}).
%
%\begin{figure}[ht]
%\centering
%\includegraphics[width=13cm]{diffusionconstant.eps}
%\caption{The diffusion constant for a quark of mass $M=1.3 GeV$ for various temperatures.
%It is easy to see that the diffusion constant becomes negative for high enough energies.
%This is more evident in (\ref{b13}) when taking the large $E/T$ limit.
%\GM{I am not sure how to interpret this physically.
%Is the medium for some reason giving energy to the quark?
%Also, which units are more appropriate?} }\label{DiffusionCon}
%\end{figure}

\section{Jet quenching parameter}\label{jq}

In this Section we discuss  the jet quenching parameter in the  class of holographic models under consideration, and we estimate its numerical
value for the concrete model with potential (\ref{dilpot}) and parameters
fixed as in \cite{gkmn3}. For the holographic computation, we will  follow \cite{Liu1,Liu2}.  There is another method available  \cite{Gubser-lan}, but we will not use it here.

The jet-quenching parameter $\hat{q}$ provides a measure of the
dissipation of the plasma and it has been associated
 to the behavior of a Wilson loop joining two light-like lines.
 We consider two light-like lines  which extend for a distance $L^{-}$ and are situated distance $L$ apart in a transverse coordinate.
Then $\hat{q}$ is given by the large $L^{+}$ behavior of the Wilson loop
\be\label{QHATDEF}
W \sim e^{-\frac{1}{4\sqrt{2}}\hat{q}L^{-}L^{2}}\;.
\ee
We consider the bulk string frame metric
\be\label{BULKMETRIC}
ds^2 = e^{2A_s(r)}\left(-f(r)dt^{2} +d\vec{x}^{2} +\frac{dr^{2}}{f(r)} \right)  \;.
\ee
To address the problem of the Wilson loop we make a change of coordinates to light cone coordinates for the boundary theory
\be\label{LIGHTCONE}
x^{+}=x_1+t\quad x^{-}= x_{1}-t
\ee
for which  the metric becomes
\be\label{LIGHTMETRIC}
ds^2 = e^{2A_s}\left(dx_{2}^{2}+dx_{3}^{2}+\frac{1}{2}(1-f)(dx_{+}^{2}+dx_{-}^2)  +(1+f)dx_{+}dx_{-}+\frac{dr^2}{f}\right)\;.
\ee
The Wilson loop in question stretches across $x_{2}$, and lies at a constant $x_{+}$,$x_{3}$.  It is convenient to choose a world-sheet gauge in which
\be\label{WORLDGAUGE}
x_{-} =\tau,\quad x_{2} =\sigma\;.
\ee
Then the action of the string stretching between the two lines is given by
\be\label{WILSACTION}
S =\frac{1}{2\pi \ls^2} \int d\sigma d\tau \sqrt{-det(g_{MN}\partial_{\alpha}X^{M}\partial_{\beta}X^{N})}
\ee
and assuming a profile of $r=r(\sigma)$ we obtain
\be\label{WILSANS}
S = \frac{L^{-}}{2\pi \ls^2} \int d x_{2} ~e^{2A_s}\sqrt{\frac{(1-f)}{2}\left(1+\frac{r'^2}{f}\right)} \;.
\ee
The integrand does not depend explicitly on $x_{2}$, so there is a conserved quantity, $c$:
\be\label{CONSQUANT}
r'\frac{\partial\mathcal{L}}{\partial r'} -\mathcal{L} ={c\over \sqrt{2}}
\ee
which leads to
\be\label{rp}
r'^2 = f\left(\frac{e^{4A_s}(1-f)}{c^2}-1 \right)\;.
\ee
\begin{figure}[h]
\centering
\includegraphics[width=10cm]{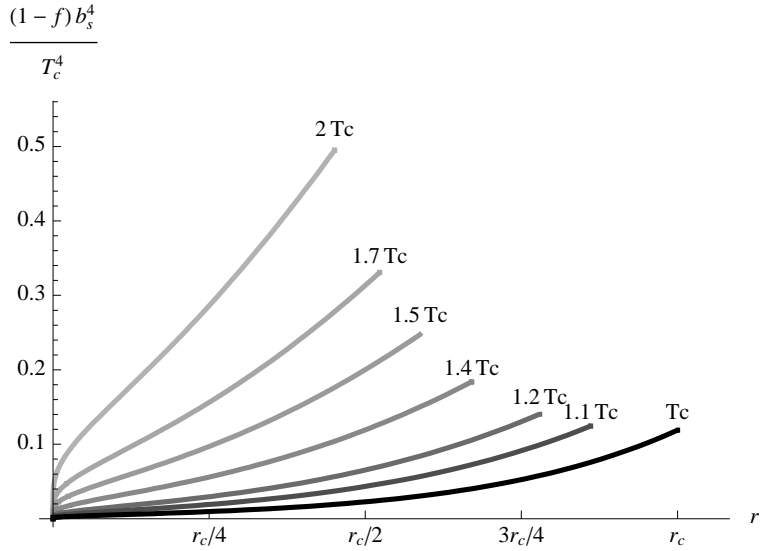}
\caption{In this figure the combination $(1-f)e^{4A_s}$ is plotted
as a function of the radial distance, for several temperatures.
The radial distance is given in units of the horizon position
$r_c$ for the black hole at the critical temperature $T_c$. All
curves stop at the corresponding horizon position.}\label{jqpic}
\end{figure}
A first assessment of this relation involves determining the zeros
and the region of positivity of the right-hand side. $f$ is always
positive and vanishes at the horizon. For the second factor we
need the asymptotics of $e^{4A_s}(1-f)$. This factor remains
positive and bounded from below in the interior and up to the
horizon. It vanishes however logarithmically near the boundary as
\be e^{4A_s}(1-f)= \pi T\ell{e^{3A(r_h)}}\left(-{1\over
\beta_0\log(\Lambda r)}\right)^{8\over 3} \left[1+{\cal
O}\left({1\over \log(\Lambda r)}\right)\right] \label{b15}\ee This
is unlike the conformal case where we obtain a constant \be
e^{4A_s}(1-f)\Big |_{\rm conformal}=(\pi T\ell)^4 \label{b16}\ee
The behavior in (\ref{b15}) is a model artifact and is analyzed in
appendix \ref{UV}.

Because of this, for fixed $c$, there is a region near the
boundary where $r'^2$ becomes negative. At this stage we will
avoid this region, by using a modified boundary at $r=\epsilon$.
We will later show that this gymnastics will be irrelevant for the
computation of the jet quenching parameter, as it involves
effectively the limit $c\to 0$.

We will place the modified boundary $r=\epsilon$ a bit inward from
the place $r=r_{min}$ where the factor
$\frac{e^{4A_s}(1-f)}{c^2}-1$ vanishes: \be
e^{4A_s(r_{min})}(1-f(r_{min}))=c^2 \label{b17}\ee Therefore we
choose $r_{min}<\epsilon$.

Then, in the range $\epsilon <r<r_h$ the factor $\frac{e^{4A_s}(1-f)}{c^2}-1$ is positive for sufficiently  small $c$.
In this same range, $r'$ vanishes only at  $r=r_h$.
This is the  true turning point of the string world-sheet.
 This is also what happens in the conformal case.
 It is also intuitively obvious that the relevant Wilson loop must sample
also the region near the horizon.

The constant $c$ is determined by the fact that the two light-like Wilson loops are a $x_{2}=L$ distance apart.
\be
\frac{L}{2} =\int_{\epsilon}^{r_{h}} \frac{c dr}{\sqrt{f(e^{4A_s}(1-f)-c^2)}}\;.
\label{b18}\ee
The denominator vanishes at the turning point. The singularity is
  integrable\footnote{Even if we choose $\epsilon=r_{min}$, the new singularity at $r=r_{min}$ is also integrable as suggested from
(\ref{b15}).}.
Therefore, as we are interested in the small $L$ region, it is obvious from the expression above that that
$c$ must also be small in the same limit.

This relation can then be expanded in powers of $c$ as
\be
\frac{L}{2c} =\int_{\epsilon}^{r_{h}} \frac{e^{-2A_s}dr}{\sqrt{f(1-f)}}+{c^2\over 2}
\int_{\epsilon}^{r_{h}} \frac{e^{-6A_s}dr}{\sqrt{f(1-f)^3}}+{\cal O}(c^4)\;.
\label{b19}\ee
Therefore to leading order in $L$
\be
c={L\over 2\int_{\epsilon}^{r_{h}} \frac{e^{-2A_s}dr}{\sqrt{f(1-f)}}}+{\cal O}(L^3)
\label{b20}\ee

We are now ready to evaluate the Nambu-Goto action of the extremal configuration we have found.
Starting from (\ref{WILSANS}), we substitute $r'$ from  (\ref{rp}), and change integration variable from $x_2 \to r$  to obtain
\be
S = \frac{2L^{-}}{2\pi \ls^2} \int_{\epsilon}^{r_h} dr ~{e^{4A_s}(1-f)\over \sqrt{2f\left(e^{4A_s}(1-f)-c^2\right)}} \;.
\label{b23}\ee

%As it always happens in asymptotically AdS spaces,  this is
%infinite, due to the infinite parts of the world-sheet near the
%boundary.
 As in \cite{Liu1,Liu2}, we
subtract from equation (\ref{b23}) the action of two free string straight worldsheets that
hang down to the horizon. To compute this action a convenient
choice of gauge is $x_{-} =\tau,\quad r =\sigma$. The action of
each sheet is \be S_{0}={L^-\over 2\pi
\ls^2}\int_{\epsilon}^{r_h}~dr~\sqrt{g_{--}g_{rr}}= {L^-\over 2\pi
\ls^2}\int_{\epsilon}^{r_h}~dr~e^{2A_s}\sqrt{1-f\over 2f}
\label{b24}\ee

The subtracted  action is  therefore:

\be\label{ACTIONSMALLC}
S_{r}= S-2S_{0} = \frac{L^{-}c^2}{2\pi \ls^2} \int_{\epsilon}^{r_h} \frac{dr}{e^{2A_{s}}\sqrt{f(1-f)}}+{\cal O}(c^4)\;,
\ee
Using now (\ref{b20}) to substitute $c$ we finally obtain
\be\label{ACTIONSMALLL}
S_{r} =\frac{L^{-} L^{2}}{8 \pi \ls^2} \frac{1}{\int_{\epsilon}^{r_h} \frac{dr}{e^{2A_{s}}\sqrt{f(1-f)}}}+{\cal O}(L^4)\;.
\ee

So far we have evaluated the relevant Wilson loop in the
fundamental representation (by using probe quarks). On the other
hand, the Wilson loop that defines  the jet-quenching parameter is
an adjoint one. We can obtain it in the large-$N_c$ limit from the
fundamental using $tr_{\rm Adjoint} =tr_{\rm Fundamental}^2$. We
finally extract the jet-quenching parameter as \be\label{WILSON}
\hat{q} = \frac{\sqrt{2}}{\pi \ls^2}\frac{1}{\int_{\epsilon}^{r_h}
\frac{dr}{e^{2A_s}\sqrt{f(1-f)}}}\;. \ee We are now ready to
remove the cutoff. As the integral appearing is now well-defined
up to the real boundary $r=0$ we may rewrite it as \be
\int_{\epsilon}^{r_{h}}
\frac{e^{-2A_s}dr}{\sqrt{f(1-f)}}=\int_{0}^{r_{h}}
\frac{e^{-2A_s}dr}{\sqrt{f(1-f)}}-I(\epsilon) \sp
I(\epsilon)=\int_0^{\epsilon} \frac{e^{-2A_s}dr}{\sqrt{f(1-f)}}
\label{b21}\ee In appendix \ref{uvinte} we obtain the small
$\epsilon$ estimate of $I(\epsilon)$ that vanishes as $\sim
\epsilon (\log\epsilon)^{4\over 3}$. We may finally
write\footnote{In practise, the previous discussion including
regularizing the UV is academic. The numerical calculation is done
with a finite cutoff where the boundary conditions for the
couplings are imposed.} \be\label{b22} \hat{q} =
\frac{\sqrt{2}}{\pi \ls^2}\frac{1}{\int_{0}^{r_h}
\frac{dr}{e^{2A_s}\sqrt{f(1-f)}}}\;. \ee

\begin{figure}[h!]
\centering
\includegraphics[width=13cm]{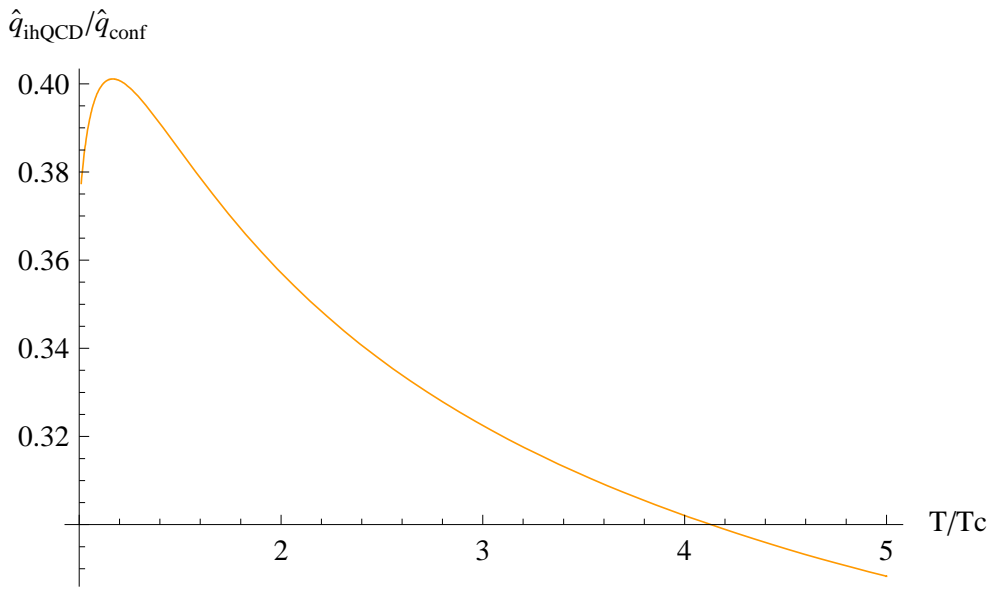}
\caption[]{In this figure the ratio of the jet quenching parameter
in our model to the jet quenching parameter in ${\cal N}=4$ is shown.
The integral present in equation (\ref{WILSON}) has been numerically calculated from
an effective cutoff at $r=r_{h}/1000$.  The jet quenching parameter in
${\mathcal N}=4$ SYM has been calculated with $\l_{'t Hooft}=5.5$.
}\label{MyFigureQhatA}
\end{figure}

\begin{figure}[h!]
\centering
\includegraphics[width=13cm]{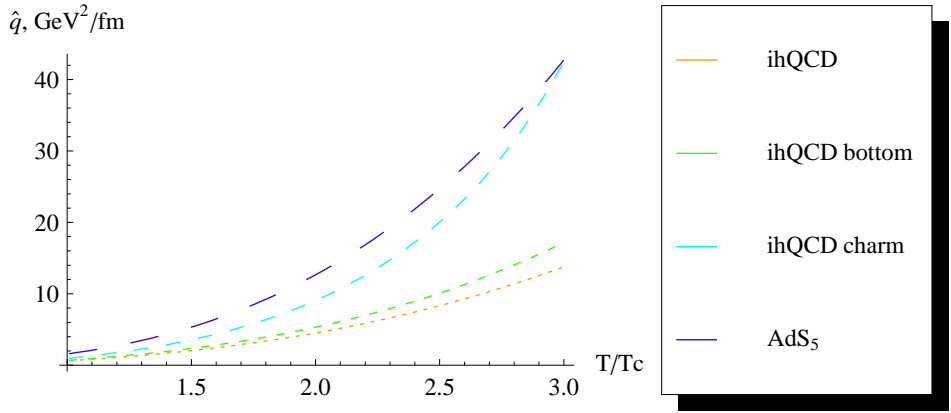}
\caption{ The jet quenching parameter $\hat{q}$ for the Improved Holographic QCD model and $\mathcal{N}=4$ SYM is shown in units of $GeV^2/fm$ for a region close to $T=T_{c}$.  The smallest dashed curve is the ihQCD result with an effective cutoff of $r_{cutoff}=r_{h}/1000$. The small dashed curve is the ihQCD result with the cutoff from the mass of the Bottom quark.  The medium dashed curve has a cutoff coming from the Charm mass and and largest dashed curve is the $\mathcal{N}=4$ SYM result.}\label{MyFigureQhatBTwo}
\end{figure}

\begin{figure}[h!]
\centering
\includegraphics[width=13cm]{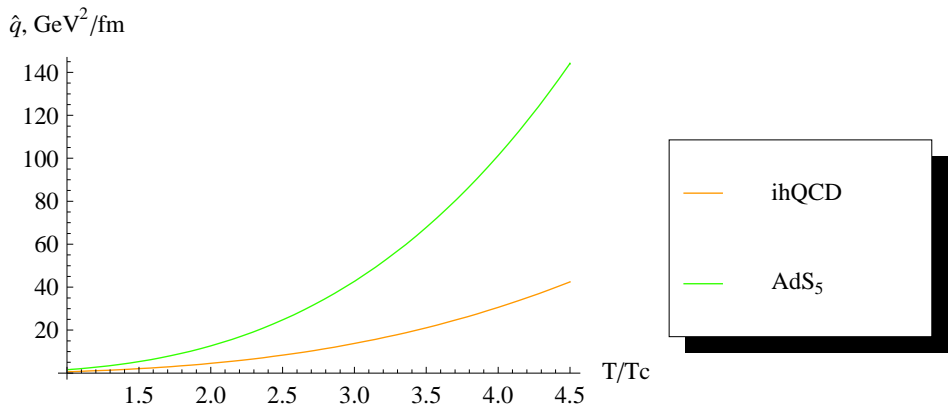}
\caption{The jet quenching parameter $\hat{q}$ for the Improved
 Holographic QCD model (lower curve) and $\mathcal{N}=4$ SYM (upper curve)
  are shown in units of $GeV^2/fm$ for temperatures up to $T=4T_c$.}\label{MyFigureQhatB}
\end{figure}

\begin{table}[h!]
\centering
\begin{tabular}{|c|c c |}
\hline
$T_{QGP},MeV$ & $\hat{q}$ $(GeV^2/fm)$  & $\hat{q}_{1}$ $(GeV^2/fm)$  \\
& (direct) & (direct)  \\
\hline
220  & -  & -    \\
\hline
250 & 0.5 & 0.6   \\
\hline
280 & 0.8 & 0.8   \\
\hline
310 & 1.1 & 1.1    \\
\hline
340 & 1.4 & 1.4  \\
\hline
370 & 1.8 & 1.8 \\
\hline
400 & 2.2 & 2.2  \\
\hline
\end{tabular}
\caption{
This table shows the jet quenching parameter $\hat{q}$ computed
 with different cutoffs for the different temperatures shown in the first column.
 The computation is done in the direct scheme.  The second column shows $\hat{q}$
 with a cutoff at $r_{cutoff}=r_{h}/1000$, where $r_{h}$ is the location of the horizon. In accordance with the conclusions of appendix  %\ref{uvinte}
 $\hat{q}$ does not change significantly as we vary the cutoff from
 $r_{h}/1000$ to $r_{h}/100$.
}\label{QHATCUTOFF}
\end{table}

\begin{table}[h!]
\centering
\begin{tabular}{|c|c c c|}
\hline
$T_{QGP},MeV$ & $\hat{q}$ $(GeV^2/fm)$  & $\hat{q}$ $(GeV^2/fm)$  &  $\hat{q}$ $(GeV^2/fm)$   \\
& (direct) & (energy) & (entropy) \\
\hline
220  & -  & 0.9  & 1.0   \\
\hline
250 & 0.5 & 1.2  & 1.3 \\
\hline
280 & 0.8 & 1.6  & 1.7  \\
\hline
310 & 1.1 & 2.1  & 2.2   \\
\hline
340 & 1.4 & 2.7 & 2.8 \\
\hline
370 & 1.8 & 3.4 & 3.4 \\
\hline
400 & 2.2 & 4.2 & 4.2 \\
\hline
\end{tabular}
\caption{This table displays the jet quenching parameter $\hat{q}$ using the three different comparison schemes. For lower temperatures the ``entropy scheme`` gives higher values.  As energy is increased the energy and entropy schemes temperatures start to coincide and there is little difference in the jet quenching parameter as well.}\label{QHATSCHEMES}
\end{table}

\begin{table}[h!]
\centering
\begin{tabular}{|c|c c c|}
\hline
$T_{QGP},MeV$ & $\hat{q}_{charm}$ $(GeV^2/fm)$  & $\hat{q}_{charm}$ $(GeV^2/fm)$  &  $\hat{q}_{charm}$ $(GeV^2/fm)$   \\
& (direct) & (energy) & (entropy) \\
\hline
220  & -  & 1.3   & 1.5 \\
\hline
250 & 0.8 & 1.8  & 2.0  \\
\hline
280 & 1.2 & 2.6  & 2.8 \\
\hline
310 & 1.7 & 3.5  & 3.6 \\
\hline
340 & 2.2 & 4.6 & 4.7 \\
\hline
370 & 2.8 & 5.9 & 6.0 \\
\hline
400 & 3.6 & 7.6 & 7.5 \\
\hline
\end{tabular}
\caption{This table displays the jet quenching parameter $\hat{q}$ using the three different comparison schemes with an effective cutoff provided by the mass of the Charm quark. Again, for lower temperatures the ``entropy scheme`` gives higher values.  As energy is increased the energy and entropy schemes temperatures start to coincide and there is little difference in the jet quenching parameter as well.  Also when the temperature approaches the quark mass the picture of the heavy quark as a hanging string collapses and results are not reliable.}\label{QHATCHARM}
\end{table}

\begin{table}[h!]
\centering
\begin{tabular}{|c|c c c|}
\hline
$T_{QGP},MeV$ & $\hat{q}_{bottom}$ $(GeV^2/fm)$  & $\hat{q}_{bottom}$ $(GeV^2/fm)$  &  $\hat{q}_{bottom}$ $(GeV^2/fm)$   \\
& (direct) & (energy) & (entropy) \\
\hline
220  & -  & 1.0   & 1.1  \\
\hline
250 & 0.6 & 1.4  & 1.5  \\
\hline
280 & 0.9 & 1.9  & 2.0 \\
\hline
310 & 1.2 & 2.5  & 2.6 \\
\hline
340 & 1.6 & 3.2 & 3.2 \\
\hline
370 & 2.0 & 4.0  & 4.0 \\
\hline
400 & 2.5 & 5.0  & 4.9 \\
\hline
\end{tabular}
\caption[]{
 This table displays the jet quenching parameter $\hat{q}$ using the three different comparison schemes with an effective cutoff provided by the mass of the Bottom quark. The results are close to the $\hat{q}$ results computed in Table  \ref{QHATSCHEMES}
since the mass of the Bottom quark is much larger than the temperatures we examine.
}\label{QHATBOTTOM}
\end{table}

From equation (\ref{b22}) we obtain, in the conformal case:
\be
\hat q_{\rm conformal}={\Gamma\left[{3\over 4}\right]\over \Gamma\left[{5\over 4}\right]}~\sqrt{2\l}~\pi^{3\over 2}T^3
\ee
The conformal value, for the median value of $\lambda=5.5$ and $T\simeq 250$ MeV
 gives $\hat q_{\rm conformal}\simeq 1.95$ GeV$^2$/fm where we used the conversion
$1$ GeV$\simeq 5$ fm$^{-1}$.

Numerical evaluation of equation (\ref{b22}) in the non-conformal IHQCD setup\footnote{In
this case, the value of $\ell_s$ appearing in equation (\ref{b22}) is fixed as explained in Section 4.}
gives us a  value of $\hat{q}$ which is lower (at a given temperature) than the
 conformal value, as shown in Figures \ref{MyFigureQhatA},
\ref{MyFigureQhatBTwo} and \ref{MyFigureQhatB}. Tables \ref{QHATCUTOFF}
to \ref{QHATBOTTOM} display the numerical values of the jet quenching
parameter at different temperatures in the experimentally relevant range,
in different temperature matching schemes.

\section{Discussion and summary}\label{DISCUSS}

In this paper we have examined several aspects associated with the physics of thermal transport phenomena in gluon plasma with
potential applications to  heavy ion collisions.
We have used as basic model for our calculations the 5D Einstein-dilaton model
 with a potential proposed in  \cite{ihqcd1,ihqcd2}.
This is a hybrid model that incorporates features coming from string theory
 as well as features originating in YM, \cite{k}.
We have also used a potential, whose two phenomenological parameters have been fit to lattice YM data, \cite{gkmn3}.
This model, named Improved Holographic QCD is expected to be a very good approximation to several aspects of YM physics.

In this context we calculated the bulk viscosity by calculating the low-frequency asymptotics of the appropriate
two-point function of the energy-momentum tensor.
We have further calculated the drag force on a heavy quark by extending the dragging string calculation done earlier
 in the context  of ${\cal N}=4$ SYM, \cite{her,gub1,tea}. Finally we have calculated the jet-quenching parameter $\hat q$ defined by
 the expectation value of a light-like Wilson loop, by adapting the calculation of \cite{Liu1,Liu2} from  ${\cal N}=4$.
 Unlike the case of ${\cal N}=4$, our calculations here are numerical as no analytical solutions are known for
 Improved Holographic QCD with the appropriate potential. We have however derived analytically various asymptotics of the results
 relevant for high energy, low velocity or high temperature.

Before discussing the results, it is appropriate at this point to take a critical look and analyze potential sources of
(systematic)  error in our calculations.

\begin{itemize}

\item Holographic models are reliable in the context of large-$N_c$ expansion of the $SU(N_c)$ gauge theory.
Therefore a priori, our results should be understood as the leading order ${\cal O}(1)$ part in the large-N$_c$ expansion.

The issue, however, is a bit more complicated by the fact the the model we are employing is semi-phenomenological
and therefore contains two phenomenological parameters (apart from the ones expected in YM) that have been fit to data in \cite{gkmn3}.
Although several lattice data are known at large-N$_c$, \cite{largen}, others are not. In particular, the detailed thermodynamics of large-N$_c$
YM is currently being calculated on the lattice, \cite{Panero}.
Therefore not all relevant input data we used have been computed at large $N_c$.

In this sense the semi-phenomenological model
we are using is positioned somewhere between N$_c$=3, YM and N$_c=\infty$ YM.
It is known so far that the difference in many observables
in the gluon sector between these two points is of the order of 5\% or less.

\item There are no dynamical flavor degrees of freedom incorporated in the model used.
In \cite{ihqcd1,ihqcd2} the incorporation of flavor branes
was described at the semi-quantitative level. We have assumed that we work in the
``quenched" approximation: the number of flavors $N_f\ll N_c$
which implies in particular that fermions loops are suppressed by a
factor of ${N_f\over N_c}\ll 1$.

The configuration of flavor branes is  expected to  involve a pair of space-filling $D_4$ and ${\overline D_4}$ per quark flavor.
These branes enter at the AdS boundary and at some point in the interior they are expected to fuse signalling chiral symmetry breaking.
The configuration in the broken phase involves a space-filling brane that folds on itself and resembles closely the branes described
in \cite{branes} using Boundary CFT.

The bare mass of the associated quarks enters as a source boundary condition on the relevant tachyon field, \cite{ckp}.
The higher the mass the stronger is the tendency on the tachyon to diverge in the IR.
In the deconfined phase we expect that branes associated to light quarks cross the BH horizon and this signals the melting of the associated mesons.
Branes associated to heavy quarks, will fuse outside the horizon, signaling  the stability of the associated mesons. These expectations are qualitative.
They  have been observed in toy models, \cite{meson} but have not been yet
calculated in a reliable extension of the present setup, \cite{pa}.

Our estimate is that the bare quark mass is related to the flavor brane position $r_m$ in the following way:
The energy of a string stretching
from $r_m$ to $r_*$, (the equilibrium position of string world-sheets) is equal to the bare quark mass, as detailed in section \ref{mass}.
This is expected to be asymptotically correct when the mass of the quark is much larger than the dynamical scale of the gauge theory,
and we therefore do not expect a large source of error for charm and bottom quarks.

\item  There are several other sources of error, that enter between
using the quantities  computed here and comparing them  to the eventual experimental data.
Most of them have been described preciously, \cite{Liu1,Liu2,her,tea,gub2}, and we do not have much more to add here.
We would like however to mention one extra important aspect:
deciding the appropriate temperature to be used in comparisons with data.
This is an issue because in  YM the deconfining transition is first order
(instead of the expected cross-over in the theory with quarks)
with a transition temperature that is about 50\% larger than in QCD.

There is therefore a non-trivial comparison to be made.
We do not know the best way to compare, but we have explored three different matchings:
taking the same temperature, the same energy density or the same entropy density.
Until a computation is made taking into account the fermionic degrees of freedom,
this choice introduces an extra systematic error in the comparison.

This ambiguity is the same one that arises when one fixes the temperature
in the holographic computations using the  ${\cal N}=4$  theory. In that  case however,
 one must also fix the ${\cal N}=4$ coupling constant, and this introduces an extra source of error. In our model this is not an issue,
since the coupling constant runs. All observable  quantities we compute are
independent of the value of the coupling at a given energy,  thus we do not need to fix an extra parameter by hand.

\item There are further limitations on the range of applicability of the drag force and
 jet-quenching calculations, that have been discussed in the literature
\cite{ct,wie}. In the drag-force calculation the velocity is limited by the position
 of the associated flavor brane. In a sense the world-sheet horizon should be kept away from
 that brane so that standard calculations of the drag force are reliable.
 The jet-quenching parameter seems valid in the opposite regime.
\end{itemize}

There is a further important issue concerning the physics of heavy quarks in the context of the QGP.
It has been argued from various points of view,\cite{tea},\cite{lan1}-\cite{iancu} that  the motion of  a heavy quark
propagating  and interacting in QGP is very similar to Brownian motion. The associated description starts with a Langevin equation
 which contains two ingredients: a classical force (the drag force) and a fluctuation force (the ``noise'') characterized by a
 a diffusion coefficient in the (late-time) Gaussian case.

 The distribution for the kinematic data then describes a Fokker-Planck equation. In the standard non-relativistic case,
 the assumption of a Maxwell equilibrium solution to the Fokker-Planck equation provides a relation between the classical
  force and the stochastic force known in the simplest cases as the Einstein relation.

  One of the relevant ingredients in the case of QGP is that the description of the Brownian motion must be relativistic.
  Relativistic Langevin evolutions have been described already in the relevant mathematical literature, \cite{math}.
  However, in the early literature, it was assumed that  the relativistic Maxwell distribution
  is an equilibrium  solution to the Fokker-Planck
  equations. This  leads to an Einstein-like relation that is problematic at high temperatures.
  This is taken as a hint that the initial assumption is false. Moreover, in the case of heavy-quark diffusion the longitudinal
   and  transverse directions behave very differently.
   A recent series of papers, \cite{Gubser-lan}-\cite{iancu} derived the Langevin-type evolution of a heavy quark in the context of AdS/CFT by
   studying the small oscillations of the trailing string solution that describes the average motion of the quark.
   In particular, it was shown in \cite{iancu} that the fluctuating force is strongly influenced by the existence of an induced 2d black-hole metric and
   an associated world-sheet horizon in the semiclassical trailing string solution, as detailed in section \ref{drag}.
   In the non-relativistic limit, this world-sheet horizon and the bulk  black-hole horizon coincide.
   The relevant diffusion coefficients are therefore computed from thermal two-point functions  for the string fluctuations.

   The analogous computation of such thermal correlators in our case, is more involved than the ${\cal N}=4$ case and will be reported  in a
   future publication. \\

Below we give a summary of our results.
%\begin{itemize}

\paragraph{Bulk Viscosity:} We have computed  the bulk viscosity by calculating the low frequency asymptotics of the appropriate stress tensor correlator
holographically. We find that the bulk viscosity rises near the phase transition but stays always below the shear viscosity.
It floats somewhat above the Buchel bound, with a coefficient of proportionality varying between 1 and 2.
Therefore it is expected to affect the elliptic flow at the small percentage level \cite{Heinz,heinz}.
Knowledge of the bulk viscosity is important in extracting the shear viscosity from the data.
This result is not in agreement with the lattice result near $T_c$. In particular
the lattice result gives a value for the viscosity that is
ten times larger.

The bulk viscosity keeps increasing in the black-hole branch below
the transition point until the large BH turns into the small BH at
a temperature $T_{min}$.  The bulk viscosity on the small BH
background is always larger than the respective one in the large
BH background.  In particular, we showed
that the T derivative of the quantity $\zeta/s$ diverges at
$T_{min}$. This is the holographic reason for the presence of a peak
in $\zeta/s$ near $T_c$. On the other hand, as it is shown in
\cite{gkmn2}, presence of $T_{min}$ (i.e. a small BH branch) is in
one-to-one correspondence with color confinement at zero T.
We arrive thus at the suggestion that in
a (large N) gauge theory that confines color at zero T, there shall
be a rise in $\zeta/s$ near $T_c$.

An important ingredient here was the value of the viscosity  asymptotically in the small BH branch. There we correlated
precisely its asymptotic value to the IR behavior of the potential. Taking also account the fact that this asymptotic
value is very close to the value of the bulk viscosity near $T_c$,
 we can derive bounds that suggest that the bulk viscosity cannot increase
a lot near $T_c$.

\paragraph{Drag Force:} The drag force we have calculated has the expected behavior.
 Although it increases with temperature, it does so slower than in ${\cal N}=4$ SYM, signaling the
effects of asymptotic freedom. This feature is at odds with the robustness observations
  for the drag force of \cite{Liu:2008tz}.

\paragraph{Diffusion Time:}Based on the drag force calculation we have computed the diffusion times for a heavy external quark.
  The numerical values we obtain are in agreement with phenomenological models \cite{lan}.
   To accommodate for the fact that our models exhibits a phase transition around $T=247$ $MeV$
   (i.e. about $30\%$ higher than in QCD), we compare our results using alternative schemes, as  proposed in \cite{gub2}.
For example, for an external Charm quark of momentum $p=10$ $GeV$ we find (in
the alternative scheme) a diffusion time of $\tau=2.6$ $fm$ at temperature $T=280$ $MeV$.
  Similarly, for a Bottom quark of the same momentum and at the same temperature we find $\tau=6.3$ $fm$.
  Generally the numbers we obtain are close to those obtained by \cite{lan} and \cite{lan2}.

\paragraph{Jet Quenching:}We have also calculated the jet quenching parameter of this model,
based on the formalism of \cite{Liu1,Liu2} by computing the appropriate light-like Wilson loop.
We find that $\hat{q}$ grows with temperature, but slower than the $T^3$ growth of $\mathcal{N}=4$ SYM result.
Again this can be attributed to the incorporation of asymptotic freedom in our model.
 Using the alternative scheme to compare with experiment
  we find that our results are close to the lower quoted values of $\hat{q}$.
   For example, for a temperature of $T=290$ MeV, which in the alternative ``energy scheme'' corresponds to a temperature of
   $T=395$ MeV in our model, we find that $\hat{q}\approx 2 GeV^2/fm$.

   However, the numbers obtained for this particular definition of jet quenching parameter seem rather low and indicate
   that this may not be the most appropriate definition in the holographic context.
   There are other ways to define $\hat q$, in particular using the fluctuations of the trailing string solution.
   This is gives a direct and more detailed input in the associated Langevin dynamics and captures the asymmetry
    between longitudinal and transverse fluctuations.
    It would be interesting to compute this, along the lines set in \cite{Gubser-lan,son,iancu} and we are currently pursuing that aim.

\vskip 2cm
\addcontentsline{toc}{section}{Note added in proof\label{note}}
\section*{Note added in proof}
Since this paper has appeared in the archive, two papers appeared that have a direct connection to some of the issues discussed here.
In reference \cite{panero2} a high precision lattice calculation of the thermodynamics was performed at various $N_c$.
The results suggest that the thermodynamic functions vary very little with $N_c$ although the phase transition becomes sharper as
$N_c$ increases. The thermodynamic functions and in particular the trace-anomaly calculated from the  Improved Holographic QCD model
\cite{gkmn3} match very well the lattice data.

In reference \cite{raja} a detailed study of the hydrodymamics with a high-viscosity regime was performed.
It was found that cavitation ensues for bulk or shear viscosity values a few times the PSS value, thus corroborating
earlier numerical evidence,   \cite{Viscos3}.
Our results indicate that cavitation (and therefore breakdown of the
hydrodynamic description) is not expected to happen in the deconfined phase of the quark gluon plasma.

\vskip 1.5cm
\addcontentsline{toc}{section}{Acknowledgments\label{ACKNOWL}}
\section*{Acknowledgements}
We would like to thank B. Bringoltz, M. Cacciari, J. Casalderrey-Solana, P. de Forcrand,
 R. Granier de Cassagnac, S. Gubser, U. Heinz, D. K. Hong, K. Kajantie, F. Karsch,
 D. Kharzeev, M. Panero, S. Pufu, F.Rocha, P. Romatchke, C. Salgado, S. J. Sin,
 C. Skenderis, A. Tseytlin and U. Wiedemann for discussions.

This work was  partially supported by  a European Union grant FP7-REGPOT-2008-1-CreteHEPCosmo-228644,
an ANR grant NT05-1-41861, a CNRS PICS grant \# 4172 and an ANR grant ANR-05-BLAN-0079-02.

Elias Kiritsis is on leave of absence from APC, Universit\'e Paris 7, (UMR du CNRS 7164).
\newpage

\appendix
\addcontentsline{toc}{section}{APPENDIX\label{app}}
\section*{APPENDIX}

\section{The scalar variables and evaluation of the bulk viscosity}\label{AppA}

To determine $\z$ we need to solve the fluctuation equation
(\ref{hw}) numerically. This requires knowledge of the background
functions $A$, $B$ and $f$ as functions of $\f$. A very convenient
reformulation of the Einstein's equations, especially when the
radial variable is taken as $\f$ is explained in section 7 of
\cite{gkmn2}, that we review here.

One can reduce the number of Einstein's equations by introducing
the following scalar variables: \be\label{psv} X(\f) =
\frac{\f'}{3A'},\qquad Y(\f) = \frac{g'}{4A'} \ee where we defined
$f=\exp(g)$. Note that $X$ and $Y$ are invariant under radial
coordinate transformations. These variables obey the following
first order equations: \bea\label{Xeq}
\frac{dX}{d\f} &=& -\frac43(1-X^2+Y)\le(1+\frac{3}{8X}\frac{d\log V}{d\f}\ri),\\
\frac{dY}{d\f} &=& -\frac{4}{3}(1-X^2+Y)\frac{Y}{X}. \label{Yeq}
\eea As shown in \cite{gkmn2}, the thermodynamics of the dual
field theory are completely determined by knowledge of $X$ and $Y$
as a function of $\f$. Roughly speaking, $Y$ is dual to the
enthalpy and $X$ to the energy of the gluon fluid.

In solving (\ref{Xeq}) and (\ref{Yeq}) one imposes the boundary
conditions at the horizon. The regularity of horizon requires
\bea\label{XYh} Y&\to& \frac{Y_h}{\f_h-\f} + {\cal O}(1),\nonumber\\
X&\to& -\frac43 Y_h + {\cal O}(\f_h-\f), \eea as $\f\to\f_h$.
Solving (\ref{Xeq}) near the horizon determines \be\label{Yh} Y_h
= \frac{9V'(\f_h)}{32V(\f_h)}. \ee

Having solved for $Y$ and $X$, one determines the metric functions
$A$ and $f$ as, \bea
A &=& A_0 + \int_{\f_0}^{\f} \frac{1}{3X}d\tilde{\f},\label{Adet}\\
g &=& \log f =  \int_{-\infty}^{\f}
\frac43\frac{Y}{X}d\tilde{\f}.\label{gdet} \eea Now, let us
compute the last metric function $B$. The metric written in the
r-frame and the $\f$-frame are: \be \label{metricrf} ds^2 =
e^{2A}(-f dt^2 + d\vec{x}^2 + \frac{dr^2}{f}) = e^{2A}(-f dt^2 +
d\vec{x}^2) + e^{2B}\frac{d\f^2}{f}. \ee Comparison determines,
\be\label{B1} B = A -\log|\frac{d\f}{dr}|.\ee In the formulation
of the scalar variables, $d\f/dr$ is given by, \be\lab{dfdr}
\frac{d\f}{dr} = -\frac{3X}{\ell}
e^{A-\frac43\int_{-\infty}^{\f}X}. \ee Thus, one finds $B$ as
\be\lab{B2} B =\frac43\int_{-\infty}^{\f}X - \log|3X|.\ee

Having found the metric functions in $X$ and $Y$ variables, one
can rewrite the fluctuation equation (\ref{hw}). There are various cancellations most notably in rewriting the
$\o$-dependent term in (\ref{hw}):  The temperature T is determined by the following equation,
(see equation (H.67) pf \cite{gkmn2}) in the scalar variables:
\be\lab{TXY}
 T=  \frac{Y(\f_0)}{\pi\ell} e^{A_0-\int_{\f_0}^{\f_h} d\f \frac{1}{X}}.
 \ee
 Now, using the equations (\ref{Yeq}), (\ref{Adet}), (\ref{gdet}) and (\ref{TXY}), the $\o$ dependent term
 can be simplified  as $$\left(\frac{wY}{3\pi T X}\right)^2 e^{-2\int_{\f}^{\f_h}\frac{1}{X}}.$$
With similar simplifications, the entire (\ref{hw}) equation can be written only in terms of $X$ and $Y$ functions:
\be\label{hwXY}
h_{11}'' = c(\f) h_{11}'+ d(\f) h_{11},\ee
where
\bea\label{cf}
c(\f) &=& \frac{1-X^2+Y}{X}\left(\frac83+\frac{3}{2X}\frac{V'}{V}\right), \\
d(\f) &=& -\frac{16Y}{9X^2}(1-X^2+Y)(1+\frac{3}{8X}\frac{V'}{V}) -
\left(\frac{\o Y}{3\pi T
X}\right)^2e^{-2\int^{\f_h}_{\f}\frac{1}{X}}\label{df}. \eea To
summarize: Given $\f_h$, one computes the functions $X$ and $Y$
from (\ref{Xeq}) and (\ref{Yeq}) and the temperature from
(\ref{TXY}). Given these data, one solves (\ref{hwXY}) numerically
(with the boundary conditions explained below (\ref{hw})).

 In passing, we note that the equation (\ref{hwXY}) can be put in a Riccati form by the change of variables $h_{11} =  \exp(\int h)$:
 $h' + h^2 = c h + d$ whose general solution can be found iff one knows a special solution. It is presumably possible to find a
 special solution for simple potentials $V$.

\section{Bulk viscosity in the limit of vanishing black-hole}\label{AppB}

Here, we fill in the details of the computation that leads
to equation (\ref{sbhlim}). This follows from (\ref{zs}) in
the high T limit on the small BH ($\l_h\to\infty$). We first show
that, the fluctuation coefficient $|c_b|$ goes to 1 in this limit.
$c_b$ is given by the value of $h_{11}(\l_h)$ that follows from
solving (\ref{hwXY}) with $\omega=0$, and the boundary condition
$h_{11}(-\infty) =1$.

In \cite{gkmn2}, it was shown that in the
$\l_h\to\infty$ limit, the functions $X$ and $Y$ simplify. In
particular $X(\l)\to X_0(\l)$ where $X_0$ corresponds to the zero
T solution and $Y(\l)\to 0$ everywhere except $\l = \l_h$. In
fact, one can show that $Y$ is proportional to a delta function
$\delta(\l-\l_h)$ in the limit $\l_h\to\infty$. Thus, from
(\ref{df}) we observe that $d(\f)$ vanishes in this limit for all
values of $\l<\l_h$. In fact it also vanishes at $\l_h$ because
the term $1+3V'/8XV$ vanishes as $\l=\l_h\to\infty$. Therefore the
fluctuation equation simplifies to \be \lab{fe1} h''_{11} (\f)=
c_0(\f) h'_{11}(\f),\,\,\,\,\, c_0(\f) = \frac{1-X_0^2+Y}{X_0}
\le(\frac83 +\frac{3V'}{2X_0 V}\ri). \ee The solution with the
aforementioned boundary condition is, \be\lab{fe2} h_{11}(\f) = 1
+ C \int_{-\infty}^{\f} dt\,\, e^{\int_{-\infty}^{t} c_0(t)}. \ee
The integration constant $C$ is determined by the second boundary
condition $h'(\f_h)=0$. On the other hand,
$c(\f)$ is positive definite in  the limit $\f\to\infty$. This is
because $V'/V$ approaches to $4/3$ whereas $X_0$ approaches to
$-1/2$. Hence the only way to obey the condition is to set $C=0$,
hence $h_{11} = 1$ for all values of $\l$ in the limit
$\l_h\to\infty$. We checked that this is indeed the case by
numerical analysis.

\section{The adiabatic approximation in scalar variables}\label{AppC}

The approximate solution explained in \cite{GubserMimick} is given
by eqs. (\ref{adb1}) and (\ref{adb2}). Using $\log s \propto A$
and (\ref{Adet}), we observe starting  from (\ref{adb1}) that the
approximation translated in scalar variables implies,
\be
\lab{C1}
X\approx X_{adb}(\f)\equiv -\frac38 \frac{V'(\f)}{V(\f)}.
\ee
 To
verify that the second equation (\ref{adb2}) leads to the same conclusion,
we may use equation (7.38) of \cite{gkmn2}:
\be
\lab{C2} \log s
- 3\log T \propto -4 \int^{\f_h} X - 3\log V(\f_h).
\ee
 On the
other hand, $\log V(\f_h)\propto -\frac83 \int^{\f_h}X$
\cite{gkmn2}. Therefore, we verify that (\ref{adb2}) also leads to
(\ref{C1}).

We compare both sides of equation (\ref{C1}) in figure
\ref{figadb3} for a large enough $\l_h$ (so that a wider range can
be compared). On this figure we also plot $X_0$ (the variable $X$
for the zero-T theory) for comparison.

\begin{figure}
 \begin{center}
\includegraphics[height=6cm,width=9cm]{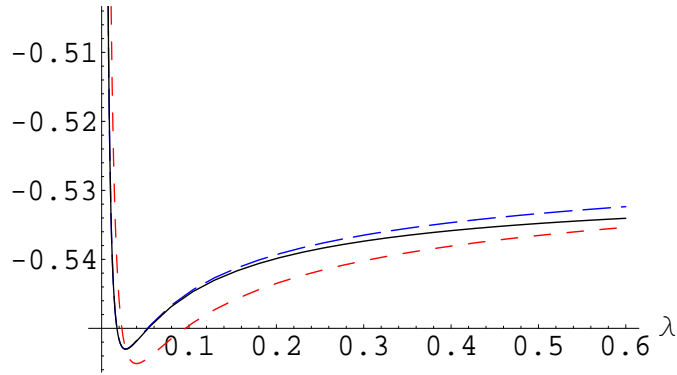}
 \end{center}
 \caption[]{Comparison of the scalar function $X$, the adiabatic approximation $X_{adb}$ and the corresponding zero-T variable $X_0$.
$\l_h=1$ is chosen to be 1. Solid (black), short-dashed(red) and
the long-dashed(blue) curves correspond to the full numerical
result $X$, the adiabatic approximation $X_{adb}$ and the zero-T
result $X_0$ respectively.} \label{figadb3}
 \end{figure}

We will now proceed to understand the approximate formula (\ref{C1})
independently. Suppose that $V'/V$ is a slowly varying function of
$\phi$. Then, we claim that we can write
\be
\lab{C3} X(\f) =
X_{adb}(\f) + \delta(\f),
\ee
 where $\delta(\f)$ is small w.r.t
$X_{adb}$ everywhere (this also  means that $\delta'(\f)$ is small
everywhere).

 Substitution of (\ref{C3}) in (\ref{Xeq})
gives\footnote{The proportionality constant is smooth and order
one. This is firstly because $X\in(-1,0)$ everywhere, and
secondly, at the point $Y$ diverges, i.e. at $\f_h$, the boundary
condition (\ref{XYh}) guarantees that the proportionality constant
is still order 1.},
\be
\lab{C4} X'_{adb} \propto \delta.
\ee
 Therefore,
the condition $X\approx X_{adb}$ is equivalent to $V'/V$ is slowly
varying with $\f$, namely the condition for the adiabatic
approximation. This argument also shows that in the limits where
$V'/V$ becomes constant, in particular near the boundary
$\f\to-\infty$, and near the singularity $\f\to+\infty$ (for
$\f_h\to\infty$), the approximation becomes exact. Figure
\ref{figadb4} supports our arguments above by numerical evidence.
Here we plot the ratio
$\le(\zeta/s(exact)-\zeta/s(adb)\ri)/(\zeta/s)(exact)$, namely the
difference between the exact (numerical) result and the adiabatic
approximation (normalized by the exact value) and the function
$|(V'/V)'(\f)|$. The latter provides the criterion for the validity
of the adiabatic approximation. The regions where both functions
become large (the region around $\l_c$) coincide, as expected from
our argumentation above. We also see that the approximation
becomes better near the UV and the IR regions.

\begin{figure}
 \begin{center}
\includegraphics[height=6cm,width=9cm]{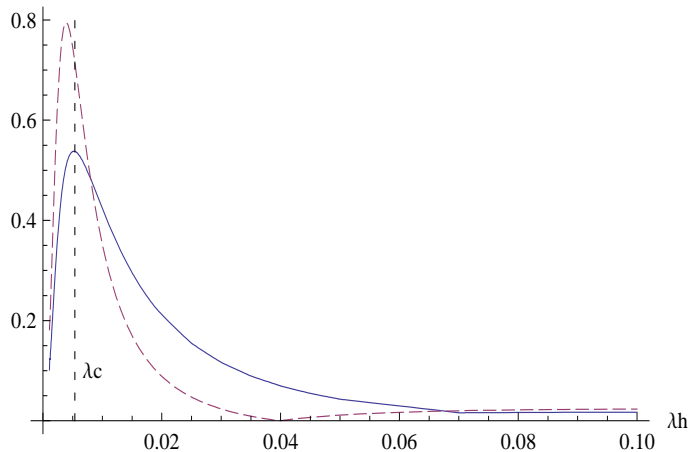}
 \end{center}
 \caption[]{Validity of the adiabatic approximation. Solid(blue) curve is the difference between the true numerical result and the adiabatic
approximation (normalized by the true value)
$\le(\zeta/s(true)-\zeta/s(adb)\ri)/(\zeta/s)(true)$, and the
dashed(red) curve is the criteria $|(V'/V)'(\f)|$ for the validity
of approximation.} \label{figadb4}
 \end{figure}

In passing let us also note the physics features that cannot be
captured by the adiabatic approximation.
 The same condition, namely that
$V'/V$ varies slowly, also leads to $X_0\big|_{adb} = -\frac38
\frac{V'(\f)}{V(\f)}$ for the zero T theory. This means $X-X_0$
vanishes in this regime. According to \cite{gkmn2}, the gluon
condensate is set to zero within this approximation. Therefore,
one cannot observe the phase transition at $T_c$ in the adiabatic
approximation. By the same reasoning, we learn that the adiabatic
approximation becomes worst when $X$ differs from $X_0$ most, i.e.
when the gluon condensate is largest, in other words in the region
near $T_c$, see also figure \ref{figadb4}.

Most equations simplify greatly with (\ref{C1}). In particular,
the coefficient $d(\f)$ of (\ref{df}) vanishes in the fluctuation
equation (\ref{hwXY}) for $\o=0$. Therefore with the same
arguments of Appendix \ref{AppB} we are lead to the conclusion $h_{11}
= 1$  in the adiabatic regime. Then, equation (\ref{zsadb}) follows
immediately.

\section{Equivalence of the axial and the $\delta\f=0$ gauges}\lab{GaugeEquivalence}

According to the standard AdS/CFT dictionary the metric
fluctuation $h_{11}=h_{22}=h_{33}$ is dual to $\half T_i^i$ {\em
in the axial gauge} $h_{5m}=0$, whereas our computation of the
bulk viscosity is carried out $\delta\f=0$ gauge, following
\cite{Gubser:2008sz}. In \cite{Gubser:2008sz}, it is shown that
the result is independent of the gauge choice, by performing a
gauge transformation between the two gauges and showing that this
does not affect the coupling of the fluctuation to the
corresponding operator to leading order near the boundary. In our
backgrounds, this issue is slightly more subtle, due to the
logarithmic corrections to the asymptotically AdS geometry.

Here we shall follow the steps in \cite{Gubser:2008sz} and prove
that indeed the gauge choice does not affect the coupling also in
our backgrounds. The metric in the $\delta\f=0$ gauge is given in
(\ref{metricrf}). Asymptotic forms of the metric functions near
the boundary read (in the $\l=\exp(\f)$ coordinate) , \bea
A(\l)&=& \frac{1}{b_0\l} + b \log(b_0\l)+\cO(\l), \label{Aasymp}\\
B(\l)&=& -\log(b_0\l)+\cO(\l), \label{Basymp}\\
f(\l)&=& 1 + \cO(e^{-4/b_0\l} (b_0\l)^{-4b}).\label{fasymp} \eea

We want to perform a gauge transformation from the gauge I to
gauge II where, \bea\label{g1} &I:& \,\, \delta\f=0,\qquad \delta
g_{\m\n}  = diag[-f e^{2A} h_{00},e^{2A} h_{11},e^{2A}
h_{22},e^{2A} h_{33},e^{2B} \frac{h_{55}}{f}],\\
\label{g2} &II:& \,\, \td{\delta\f}=\z^5\neq 0,\qquad \delta
\td{g_{\m\n}} = diag[-f e^{2A} \td{h}_{00},e^{2A}
\td{h}_{11},e^{2A} \td{h}_{22},e^{2A} \td{h}_{33},0]. \eea Under
an infinitesimal gauge transformation $\z_{\m}$, the metric
functions and the dilaton transform as \be\lab{gtransf} \td{\d\f}
= \d\f + \z^{\m}\6_{\m}\f, \quad \d \td{g}_{\m\n} = \d g_{\m\n} +
\del_{\m}\z_{\n} + \del_{\n}\z_{\m}.\ee The symmetries of the
problem dictate that $\z^1=\z^2=\z^3=0$ and $\z^0 = \z^0(t,\f)$,
$\z^5 = \z^5(t,\f)$.

When applied to (\ref{g1}) in order to get (\ref{g2}) these
transformations reduce to the following equations. The dilaton equation
yield $\td{\d\f} = \z^5$ and the 55, 50, 00 and 11 components of
the second equation in (\ref{gtransf}) respectively produce
\cite{Gubser:2008sz}: \bea {}&&\z^{5'} +
(B'-\frac{f'}{2f})\z^5 + \frac{h_{55}}{2}=0,\lab{tr55}\\
{}&&\z^{0'} - \frac{e^{2(B-A)}}{f^2}\dot{\z}^5=0,\lab{tr50}\\
{}&& \td{h}_{11} - h_{11} - 2A'\z^5 = 0,\lab{tr11}\\
{}&& \td{h}_{00} - h_{00} - (2A'+\frac{f'}{f})\z^5 - 2\z^{0'} =
0,\lab{tr00} \eea where prime and dot denotes derivation w.r.t.
$\f$ and time, respectively.

We shall assume an oscillatory form for the t-dependence of the fluctuations, $\delta x(t,\f) = e^{-i w t} x(\f)$ and
use the same symbol to denote the $\f$-dependent piece, with a slight abuse of notation. One can easily see that
the following computation goes through with no change for a more general t-dependence.
Using the asymptotic forms of the metric functions above, one finds
an approximate solution to (\ref{tr55}) as,
\be\label{sol55}
 \z^5(\f) \approx \l\le[ c_1 -\half\int^{\l}_{0} \frac{d\td{\l}}{\td{\l}^2} h_{55}(\td{\l})\ri].
\ee
In order to determine the asymptotic behavior, one needs to determine $h_{55}$ near the boundary. This can be done
by using the fluctuation equation for $h_{55}$ ( see \cite{Gubser:2008sz}),
\be\lab{H55fluc}
h_{55} = \frac{1}{A'}\le( h_{11}'-\frac{f'}{2f} h_{11}\ri),
\ee
where $h_{11}$ is the solution to (\ref{hw}). Again, using the asymptotic forms of the metric functions above,
one finds that the solution to (\ref{hw}) near boundary  (with the boundary condition $h_{11}\to 1$) reads,
\bea
h_{11}&\to& 1+ c \l^{-2b}e^{-\frac{2}{b_0\l}}, \qquad \o\neq 0,\lab{h11sol1}\\
h_{11}&\to& 1+ c' \l^{-4b-1}e^{-\frac{4}{b_0\l}}, \qquad
\o=0,\lab{h11sol2} \eea where $c$, $c'$ are some integration
constants. Using these in (\ref{H55fluc}) one finds that $h_{55} =
\cO(\l^{-2b}e^{-\frac{2}{b_0\l}})$  for $\o\neq 0$ and
$\cO(\l^{-4b-1}e^{-\frac{4}{b_0\l}})$ for $\o = 0$. Finally,
substituting this in (\ref{sol55}) gives (for $\o\neq 0$) ,
\be\lab{sol552} \z^5 = \l \le[ c_1+
\cO(\l^{-2b-1}e^{-\frac{2}{b_0\l}})\ri]. \ee We see that the
leading term goes as $-1/\log r$ whereas the sub-leading term is
suppressed as $\cO(r^2)$ as $r\to 0$ at the boundary. Thus we can
safely ignore the inhomogeneous contribution in (\ref{sol55}) and
take $\z^5 \approx c_1 \l$. Using this and the asymptotics of the
metric functions  above in (\ref{tr50}) now gives, \be\lab{sol50}
\z^0 \approx c_2 \l^{-2b}e^{-\frac{2}{b_0\l}} e^{-i wt} \ee
Finally, using all the above, one solves (\ref{tr00}) and
(\ref{tr11}) as (stripping off the t-dependence), \be\lab{h1100}
\td{h}_{00} = h_{00} - \frac{2c_1}{b_0} +
\cO(\l^{-2b-1}e^{-\frac{2}{b_0\l}}), \quad \td{h}_{11} = h_{11} -
\frac{2c_1}{b_0} + \cO(\l^{-2b-1}e^{-\frac{2}{b_0\l}}). \ee

The operator that is dual to $\z^5=\delta\f$ is $\cO = tr\, F^2/(4\l)$ (\cite{gkmn2}) . Thus, from (\ref{sol552}) and (\ref{h1100}) we find that the
fluctuation of the Lagrangian that is proportional to $c_1$ is
\be\lab{lagc1}
\delta_{c_1}{\cal L} = -\frac{1}{b_0}T^{\m}_{\m} + \frac14 tr\, F^2.
\ee
Expanding the dilatation Ward identity, $T_{\m}^{\m} = \frac{\beta(\l)}{4\l^2}tr \, F^2$ to leading order in $\l$ we see that, (\ref{lagc1}) vanishes.
This proves that the fluctuation of the metric function $h_{11}$ couples to $\half T_i^i$ both in the axial gauge and in the $\delta \f=0$ gauge.

\section{UV subtleties}\label{UV}

In a theory where the dilaton is non-trivial, there is a relevant question to be asked. In which frame is the metric asymptotically AdS?
This question is void in the dual of ${\cal N}=4$ SYM where the dilaton is trivial but not in Improved Holographic QCD where the gauge coupling is a function of the holographic
coordinate.
In \cite{k} by analyzing the structure of string higher-curvature corrections, it was shown that it is consistent with the equations of motion
that
 the string-frame metric is asymptotically AdS. This is required in order for the background solution to have the correct structure
 and QCD perturbation theory to emerge.

By approximating the string theory dual to QCD by a two derivative theory and a dilaton potential as it was proposed in \cite{ihqcd1,ihqcd2}
this property cannot be maintained. It is not easy to see that the only option that
can be implemented in the UV is an asymptotically AdS metric in the Einstein frame
instead. This has as a result a few ``stray logs" in several quantities that are calculated from the world-sheet action (instead of the bulk effective action).
One of them is the short distance inter-quark potential calculated in \cite{zeng}
which is\footnote{Interestingly, it was  argued in \cite{zeng} that this fits better Quarkonium data than the Cornell
potential.} $V(r)\sim {(\log (r\lambda))^{4\over 3}\over r}$.

Other similar cases appear in this paper, in the two observables that involve the string world-sheet action.
The first is the drag-force calculation. The effect of these logs appears both as the energy of the string end-point becomes asymptotically large
($v\to 1$) or when the temperature becomes large $T\to\infty$.
One example is the ultra-relativistic  diffusion time (\ref{b11}) that we reproduced here
\be
\lim_{p\to\infty}~\tau=M_q~{\ls^2\over \ell^2} \sqrt{4N_c^2\over 45 ~T s(T)}
\left({b_0\over 4}\log{p^2\over M_q^2}\right)^{4\over 3}+\cdots
\label{b111}\ee
The logarithmic fact is due to a  factor of $\l^{-{4\over 3}}$.
Another example is the large $T$ asymptotics of the non-relativistic diffusion time in (\ref{b12}).
It gives exactly the conformal result modulo again a factor of $\l^{-{4\over 3}}$.

A similar effect appears in the jet-quenching calculation in section \ref{jq}.
Indeed, in (\ref{b22}), the combination $e^{4A_s}(1-f)$ vanishes logarithmically in the UV instead of asymptoting to a constant value.
This occurrence perturbs the structure of the Wilson loop configuration near the boundary,
but as shown there does not affect the calculation of the jet-quenching parameter.

A perturbative calculation at NLO of the diffusion time gives, \cite{moore}
\be
{1\over \tau_{\rm pQCD}}={8\pi T^2\over 3M}\alpha_s^2\left[-\log g+0.07428+1.8869 g\right]
\ee
while for large $N_c$ ${\cal N}=4$ SYM it is obtained
\be
{1\over \tau_{\rm psYM}}={\l^2 T^2\over 12\pi M}\left[\log {1\over \sqrt{\l}}+0.4304+0.801 \sqrt{\l}\right]
\ee
Such perturbative asymptotics are not visible in the NG action.

There is however an important issue here: when and where we can trust the standard Nambu-Goto world-sheet action.
The structure of the vacuum solution near the boundary, advocated in \cite{k},  suggests that since curvatures are high in that region,
care is needed when using the NG action in that regime.
On the other hand, the UV behavior remains qualitatively correct although  in its details it may be revisable.

\section{The UV asymptotics of the integral (5.21) }\label{uvinte}

We now turn to estimating the integral
\be
I(\epsilon)\equiv \int_0^{\epsilon}{dr\over e^{2A_s}\sqrt{f(1-f)}}
=\int_0^{\epsilon}{\l^{-{4\over 3}}~dr\over e^{2A}\sqrt{f(1-f)}}
\label{e1}\ee
used in section \ref{jq} when $\epsilon\to 0$.

Near $r=0$ in the Einstein frame, \cite{ihqcd1},
\be
f(1-f)\simeq \pi T{e^{3A(r_h)}\over \ell^3}r^4\left[1+{\cal O}\left({1\over \log(\Lambda r)}\right)\right]
\label{e2}\ee
\be
\l\simeq -{1\over b_0\log(\Lambda r)}+{\cal O}\left({\log\log(\Lambda r)\over \log^2(\Lambda r)}\right)\sp
e^A\simeq {\ell\over r}\left[1+{\cal O}\left({1\over \log(\Lambda r)}\right)\right]
\label{e3}\ee
Using these relations we obtain
\be
I(\epsilon)=\int_0^{\epsilon}{\l^{-{4\over 3}}~dr\over e^{2A}\sqrt{f(1-f)}}\simeq
 {b_0^3\over \sqrt{\pi T~e^{3A(r_h)}\ell}}\int_0^{\epsilon} (-\log(\Lambda r))^{4\over 3}
\left[1+{\cal O}\left({\log\log(\Lambda r)\over \log(\Lambda
r)}\right)\right]dr
\label{e4}\ee
 changing variables to $u=-\log(\Lambda r)$
we obtain
\be
 I(\epsilon)={b_0^3\over \Lambda\sqrt{\pi
T~e^{3A(r_h)}\ell}}\int_{-\log(\Lambda \epsilon)}^{\infty}du~u^{4\over
3} e^{-u} \left[1+{\cal O}\left({\log(u)\over u}\right)\right]
\label{e5}\ee
We now use
\be
 \int_{-\log(\Lambda \epsilon)}^{\infty}du~u^{4\over 3}
e^{-u}=\Gamma\left[{7\over 3},-\log(\Lambda \epsilon)\right]
\simeq \left[-\log(\Lambda \epsilon)\right]^{4\over 3}~\Lambda
\epsilon+
\label{e6}\ee
to finally obtain
 \be I(\epsilon)={b_0^3\over \sqrt{\pi
Tb^3(r_h)\ell}}\left[-\log(\Lambda \epsilon)\right]^{4\over 3}~
\epsilon\left[1+ {\cal O}\left({\log\log(\Lambda \epsilon)\over
\log(\Lambda \epsilon)}\right)\right]
\label{e7}\ee
valid as $\epsilon\to 0$.

\newpage

\addcontentsline{toc}{section}{References\label{refs}}

\end{document}